\documentclass{amsart}
\RequirePackage{fix-cm}
\usepackage{amsmath}
\usepackage{ulem}
\usepackage{amssymb}
\usepackage{bbold}
\usepackage{breakcites}
\usepackage[margin=1.2in]{geometry}
\usepackage{multirow}
\newcommand{\Earth}{\textrm{E}}
\newcommand{\Moon}{\textrm{M}}
\newcommand{\Sun}{\textrm{S}}
\newcommand{\F}{\textrm{F}}

\newcommand{\norm}[1]{\left\lVert#1\right\rVert}
\newcommand{\ie}{\textit{i.e.,\ }}
\newcommand{\eg}{\textit{e.g.,\ }}
\newcommand{\dd}{\textrm{d}}

\newcommand{\DOF}{DoF\,}
\usepackage{rotating}

\usepackage{hyperref}

\begin{document}

\title[Secular evolution of Molniya semi-major axis]{Dynamical properties of the Molniya satellite constellation: long-term evolution of the semi-major axis}

\author{J.\,Daquin         
	\and
	E.M.\,Alessi \and
	J.\,O'Leary \and
	A.\,Lemaitre \and
	A.\,Buzzoni
}
\address{Department of Mathematics (naXys), $61$ Avenue de Bruxelles, $5000$, Namur, Belgium}
\email{jerome.daquin@unamur.be}

\address{Instituto  di  Matematica  Applicata  e  Tecnologie  Informatiche  ``Enrico  Magenes'',  Consiglio Nazionale delle Ricerche, Via Alfonso Corti 12, $20133$ Milano, Italy \\
	Istituto  di  Fisica  Applicata  ``Nello  Carrara'',  Consiglio  Nazionale  delle  Ricerche,  Via Madonna del Piano 10, $50019$ Sesto Fiorentino, Italy}

\address{EOS Space Systems Pty. Ltd., Lot Fourteen, McEwin Building. North Terrace, Adelaide, SA 5000, Australia}

\address{Department of Mathematics (naXys), 61 Avenue de Bruxelles, $5000$, Namur, Belgium}

\address{INAF-OAS, Osservatorio di Astrofisica e Scienza del Spazio, Via P. Gobetti $93/3$, $40129$ Bologna, Italy }

\keywords{
Molniya orbit; Tesseral resonance; Lunisolar resonance; Manifolds; Fast Lyapunov Indicator; Space Situational Awareness
}
\date{\today}

%\maketitle
%\tableofcontents
%\subjclass[2010]{Primary }
%\keywords{medium Earth orbits - numerical integration - keyword 3}
\date{\today}

\maketitle

\begin{abstract}
	We describe the phase space structures related to the semi-major axis of Molniya-like satellites subject to tesseral and lunisolar resonances.  In particular,  the questions answered in this contribution are: (i) we   study the indirect interplay of  the critical inclination resonance on the semi-geosynchronous resonance  using a hierarchy of more realistic dynamical systems,  thus discussing the dynamics beyond the integrable approximation. By introducing \textit{ad hoc} tractable models averaged over  fast angles, (ii) we numerically  demarcate the hyperbolic structures organising the long-term dynamics  via Fast Lyapunov Indicators cartography. Based on the publicly available two-line elements space orbital data, (iii) we identify two satellites, namely Molniya 1-69 and Molniya 1-87,  displaying fingerprints  consistent with the dynamics associated to the hyperbolic set. Finally, (iv) the computations of their associated dynamical maps highlight that the spacecraft are trapped within the hyperbolic tangle. This research therefore reports evidence of actual artificial satellites in the near-Earth environment whose dynamics
	are ruled by manifolds and resonant mechanisms. The tools, formalism and methodologies we present are exportable to other region of space subject to similar commensurabilities as the geosynchronous region.	
\end{abstract}

%=======
\section{Introduction}
%=======
The present manuscript is part of a recent series of papers \cite{eAl21,aBu20} dedicated to
astrodynamical properties of Molniya spacecraft.  
It is well-known that the Molniya orbit provides a  valuable
dynamical alternative to the geosynchronous orbit, suitable for
communication satellites to deliver a service in high-latitude
countries, as it is actually the case for Russia.
In the present contribution, we focus  on the 
decadal evolution of the semi-major axis.   
We approach the problem by studying the long-term and  drag-free motion of a test-particle subject to the non-spherical geometry of the Earth and third-body perturbations due to the Sun and the Moon.
The metrical Keplerian  characteristic values of the Molniya-class, semi-major axis $a$ (expressed in units of Earth radius, $r_{\Earth}$), eccentricity $e$ and inclination $i$, considered in this work are  
\begin{align}
\oe_{\mathcal{M}}
= 
(a_{\mathcal{M}},e_{\mathcal{M}},i_{\mathcal{M}})
\sim
(4.16\,r_{\Earth},0.7,63.4^{\circ}).
\end{align}

To follow this goal, the zonal geopotential terms are  first restricted to the second degree $J_{2}$ term. 
Molniya satellites have a mean motion close to $2$ revolutions per day and thus are subject to a $2:1$ resonant commensurability with the Earth's rotation rate (semi-synchronous orbits). Therefore, $12$-hour resonant terms of the geopotential need to be taken into account to model the dynamics. The resonant terms are algebraically  computed up to the $4$th degree and order.   
Being interested in  long-term dynamics,  we deal with the various Hamiltonian contributions averaged over the fast variables, leading to the so-called \textit{secular} dynamics.
The fast timescales are connected to the mean anomaly of the test-particle and the Moon  and the Sun, denoted respectively $M, M_{\Moon}, M_{\Sun}$.  
The averaged contributions are introduced as the direct computation of the integral with respect to the fast variables. For the zonal contribution, this averaging is performed in closed form with respect of the eccentricity. The quadrupolar lunisolar perturbations, depending respectively on $M$ and  $M_{\Moon}$ or $M$ and  $M_{\Sun}$ are  doubly averaged, also in closed form with respect of the eccentricity. 
For the resonant contribution of the geopotential, the averaging requires some extra care. First, the averaging is not performed in 
closed form over the eccentricity. Instead, we 
employ a truncated series expansion, which, considering the highly eccentric nature of the orbit, is given to $4$th order in the eccentricity.  Second, the averaging is not performed over the variable $M$ directly, as it would not take into account accurately the resonant dynamics. Instead, this step  calls for the introduction of new slow/fast variables taking into account the very resonant nature of the problem \cite{vBe12}. Once those variables are recognised and introduced explicitly, the averaged contribution is obtained in the usual way, \ie by averaging over the (new) fast variables. \\

Molniya, Raduga, Gorizont and Ekran are Russian communication satellites inherited from the Soviet era. Molniya is the Russian word for lightning, thus, given their interesting dynamical framework, are aptly named. The framework for the Russian communication satellite constellation was first presented by Bill Hilton in the British Interplanetary Society during the years 1959-60 \cite{bHa07} who suggested utilizing highly inclined, highly eccentricity orbits for communication systems for high latitude regions. The Molniya constellation operates on a unique orbital configuration, which, is not exploited by any other type of satellite system. While typical communication satellites operate on a $24$h geosynchronous equatorial orbit, the high latitude of many Russian areas poses a problem for radio frequency transmissions from equatorially orbiting spacecraft. For example, the Russian republic spans a range of 
$40$ degrees in latitude from North to South, with the northernmost point being located at $80$ degrees North.  The solution to the satellite communication problem for high latitude regions is the Molniya orbital regime. Operating on a highly inclined and eccentric $12$ hour orbit, the Molniya spacecraft reach geosynchronous altitude at apogee, providing access to Russian areas for over $8$ hours per orbit and reaches approximately $600$ km at perigee at much greater orbital velocities. Since their inception, over $160$ Molniya spacecraft have been launched which have provided a platform for research on their unique dynamical framework for nearly $60$ years.
Molniya orbits gather two distinct resonant phenomena\footnote{
	The force model we employed is discussed in more details in Appendix \ref{app:ForceModel}. The resonant argument $\omega$ which appears in the expansion of the lunisolar Hamiltonian also appears in higher geopotential zonal terms. In this sense, Molniya orbits gather more than $2$ resonant phenomena, being affected by zonal, tesseral  and third-body resonances.
	Nevertheless, the effects of higher zonal terms
	on the semi-major dynamics are negligible for our study and timescale of interests as we will  demonstrate later.}, with quite distinct timescales, 
giving rise to interesting qualitative dynamical behaviour. Firstly, as we mentioned, they are affected by a $2:1$ geopotential resonance. Secondly, their inclination close to the critical inclination value of $ 63.4^{\circ}$  place them near a so-called ``inclination dependent only'' lunisolar resonance \cite{sHu80}. Whilst the first affects the semi-major axis of the orbit on a yearly timescale, the lunisolar effect manifests primarily on the eccentricity of the orbit, which exhibits large oscillations on a much longer timescale. These pulsations contribute to modulate the (no-longer constant) coefficients of the tesseral problem; henceforth a coupling and indirect interplay between the two resonances might happen. The seminal contributions regarding the tesseral and lunisolar problems are gathered in \cite{fDe93,fDe93-2}, and in the PhD work of T.\,Ely \cite{tEl96}, later extended to full papers \cite{tEl97,tEl00}. F.\,Delhaise, J.\,Henrad and A.\,Morbidelli \cite{fDe93,fDe93-2} have focused their study on the eccentricity, inclination and argument of perigee, without paying attention to the behaviour of the semi-major axis. T.\,Ely \cite{tEl96} connected the resonant problem with large-scale chaos affecting the semi-major axis, including the disturbing effects of the lunisolar perturbation, but for  orbital parameters which differ
quite significantly
from Molniya orbit (in fact, he considered either moderate inclined orbits with $i \sim 20^{\circ}$ or inclinations in the vicinity of the $2g+h$ lunisolar resonance, \ie at $i \sim 56^{\circ}$). Thus, the secular dynamics of Molniya semi-major axis remains partially unexplored.  \\

The first contribution of this paper is to discuss the dynamics of the semi-major axis beyond the integrable picture. For this task, we rely on classical tools from nonlinear dynamics  to portray the dynamical structures organising the long-term dynamics (Poincar\'e section, sections of finite-time variational indicators).  The chaotic nature of eccentric and inclined orbits subject to tesseral resonances, often explained through an overlap of nearby resonances \cite{bCh79}, has been known for some time in the context of tesseral resonances \cite{aCe14,jDa15,tEl96}. Nevertheless, as we will highlight, the extent of chaos affecting the semi-major axis phase space for Molniya satellites is much more limited in the range of $i \sim 63^{\circ}$ compared to the previously studied range of inclinations. In fact, large connected chaotic seas are absent from the dynamics. Yet, hyperbolic orbits still exist and surround the unperturbed separatrix as we will show.   

The second contribution of this paper is to reveal  the precise effects of this coupling on the dynamics of the semi-major axis. This is achieved  via the introduction of several dynamical systems, aiming at isolating gradually the various effects and couplings. 
The driving principle is to introduce basic dynamical models, with the lowest number of degree-of-freedom (DoF) possible, which still encapsulate the physics and long-term qualitative features of the dynamics. 
Molniya orbits have also received attentions in \cite{tlZh15,tlZh14}, but predominantly oriented towards the description of the long-term evolution of the eccentricity. The authors have built simplified secular dynamical models, in the same spirit as ``isolating'' the building blocks of the dynamics and reconstructed the qualitative features  of the eccentricity, inclination and argument of perigee observables. A few model generated orbits have been compared to  the publicly available two-line element (TLE)  datasets (see, \eg \cite{mCa14,dVa01})\footnote{Available at
	\href{https://www.space-track.org}{space-track.org}.}. We underline that our contribution is paying particular  attention to the orbit of Molniya 1-69\footnote{North American Aerospace Defense Command (NORAD) satellite catalog number 
	$17078$, COSPAR ID 	1986-089A.}, left untouched in  a previous study, as being ``in the vicinity of the separatrix'' \cite{tlZh14}. 
TLEs remain mainly the sole reservoir of orbital data.  The TLEs result from observational measures, coupled with an orbit determination process and  numerical propagations performed with simplified theories of motion. In this respect, they form rather pseudo-observations instead of ``pure'' observational data. Approximately every $8$ hours, the unclassified TLEs are released publicly. Molniya spacecraft have been tracked since the mid-70's, thus providing a sufficient long-time interval of TLEs to appreciate secular effects acting on the semi-major axis.

The third and final contribution of this paper is the clear connection of the dynamics of two satellites,  Molniya 1-69 and  Molniya 1-87\footnote{NORAD ID $22949$, COSPAR ID 1993-079A.}, with the fingerprints of the dynamics associated to the hyperbolic set. This last point sheds some light  of the relevance of secular dynamical approaches and  toolboxes for the field of space situational awareness and the continuing increasing space traffic. The patterns of the orbital semi-major axis time-series (extracted from the corpus of TLEs, more details will be presented in the subsequent) of the two aforementioned satellites are convincingly approached under this umbrella.    

The paper is organised as follows: 
\begin{itemize}
	\item In section \ref{sec:geo}, based on the Earth-only disturbing potential,
	a secular model is termed. A resonant integrable system  is formulated from which analytical quantitative estimates (width of the resonance, characteristic timescales) are extracted.  This integrable picture is altered by a  multiplet of resonances producing a separatrix splitting phenomena, responsible for the apparition of a chaotic layer in the phase space. 
	For Molniya parameters, the overlap of resonances is complete.  
	The corresponding 2-\DOF Hamiltonian and its phase space is described via Poincar\'e sections. 
	%The chaotic layer leads to  intermittency phenomena of the semi-major axis.  
	%Taking advantage of the $2$-\DOF nature of the system and using finite-time variational chaos indicators,  the geometry of the chaotic layer as a function of the initial argument of perigee of the satellite is predicted. 
	%This point is useful in practice to relate the location of a specific satellite with respect to the hyperbolic structures of the phase space.   
	%The precise geometry of the chaotic layer is revealed using numerical finite-time  variational chaos indicator. 
	\item In section \ref{sec:LS}, we introduce two models including lunisolar perturbations to overcome the
	limitations of the Earth-potential only based model. From these models, the effects of the lunisolar perturbations on the tesseral problem are studied. We use dynamical indicators to portray the phase space structures and reveal the hyperbolic set affecting the semi-major axis. The dynamics of the hyperbolic set is studied.
	\item In section \ref{sec:TLE}, after providing more information about the TLEs datasets, we
	connect the dynamics of the  data for  satellites Molniya 1-69 and Molniya 1-87 with
	the dynamics 
	of the hyperbolic set. 
	Relying on our understanding of the underlying dynamics, we extract specific epochs and orbital parameters of the TLEs that spot the satellites within the hyperbolic tangle when computing their respective  dynamical maps. 
\end{itemize} 
We close the paper by summarising our  conclusions.

%=======
\section{The secular and geopotential based Hamiltonian}\label{sec:geo}
%=======
We present our steps and assumptions to recover a relevant secular Hamiltonian model for $12$-hour orbits based on the geopotential only and we describe the associated dynamics. \\

The disturbing 
potential of the Earth, in an Earth-centered and Earth-fixed frame, admits the
following expansion  \cite{wKa66}
\begin{align}\label{eq:PertPotential}
V(r,\phi,\lambda)=
V_{\mathcal{Z}}(r,\phi) + V_{\mathcal{T}}(r,\phi,\lambda),
\end{align} 
with the zonal and tesseral parts respectively given by 
\begin{align}
\left\{
\begin{aligned}
&V_{\mathcal{Z}}(r,\phi)=\frac{\mu}{r} \sum_{l \ge 2} 
\Big(\frac{r_{\Earth}}{r}\Big)^{l}
J_{l,0}P_{l,0} \sin \phi, \\
&V_{\mathcal{T}}(r,\phi,\lambda)=-\frac{\mu}{r}
\sum_{l \ge 2}  
\sum_{m=1}^{l} 
\Big(\frac{r_{\Earth}}{r}\Big)^{l}
\Big(c_{l,m} \cos m\lambda + s_{l,m} \sin m\lambda\Big)P_{l,m}(\sin \phi),
\end{aligned}
\right.
\end{align} 
where the vector $(r,\phi,\lambda)$ denote the spherical coordinates (respectively radius, latitude and longitude), $r_{\Earth}$ denotes the mean Earth's radius, $\mu$ the gravitational parameter of the Earth. 
The $P_{l,m}$ are the Legendre polynomials of degree $l$ and order $m$. The coefficients $c_{l,m}$ and $s_{l,m}$ are the harmonic 
coefficients describing Earth's gravity field where we  denoted classically $J_{l,0}=-c_{l,0}$.    
Throughout this paper,  we denote the Keplerian orbital elements in the usual way as $(a,e,i,\omega,\Omega,M)$ where $a$ denotes the semi-major axis, $e$ the eccentricity, $i$ the inclination, $\omega$ the argument of perigee, $\Omega$ the longitude of the ascending node and $M$ the mean anomaly.

%======
\subsection{The secular zonal part}\label{subsec:zonal}
%======
The zonal part 
is dominated by its quadrupole $(l=2)$ term and we therefore truncate  $V_{\mathcal{Z}}$ to $l=2$.    
Being interested by secular properties, 
the $M$-average of the $J_{2}$ part, defining the secular $J_{2}$ contribution, is  computed (in closed form over the eccentricity) using  
the differential relationship
\begin{align}
\dd M = \frac{r^{2}}{a\sqrt{1-e^{2}}}\dd f,
\end{align}
together with the formula 
$r=a(1-e^{2})/(1+e \cos f)$ (see \eg \cite{wKa66}): 
\begin{align}\label{eq:AvJ2}
\bar{V}_{J_{2}}=\frac{1}{2\pi}\int_{0}^{2\pi}V_{J_{2}} \, \dd M
=
\frac{1}{2\pi}\int_{0}^{2\pi} \frac{r^2}{a^2\sqrt{1-e^2}}V_{J_{2}} \, \dd f.
\end{align}
The classical final expression (\ref{eq:AvJ2}), expressed in terms of the orbital elements, reads
\begin{align}
\bar{V}_{J_{2}}=\frac{\mu r_{\Earth}^{2}J_{2}}
{4a^{3}}(1-e^{2})^{-3/2}(3 \sin^{2} i-2).
\end{align}
Secular expressions of higher order terms and their dynamical effects are  
discussed
in Appendix \ref{app:ForceModel}. We note, for the remainder of the manuscript, we drop  bars over averaged quantities, bearing in mind that we are dealing with secular functions in this study. 

%======
\subsection{The secular resonant tesseral part for $12$-hour orbits}\label{subsec:tess}
%======
To compute the resonant secular contribution of the tesseral part 
\begin{align}\label{eq:VTess}
V_{\mathcal{T}}=
\sum_{l \ge 2} \mathcal{T}_{l},
\end{align}
with 
\begin{align}
\mathcal{T}_{l}= \sum_{m=1}^{l}-
\frac{\mu}{r} \Big(\frac{r_{\Earth}}{r}\Big)^{l}
\Big(c_{l,m} \cos m\lambda + s_{l,m} \sin m\lambda\Big)P_{l,m}(\sin \phi),
\end{align}
we first express it in terms of the orbital elements using a series of formal substitutions.  The spherical coordinates are related to the orbital elements by:

\begin{align}
\left\{
\begin{aligned}
&\cos(\alpha-\Omega)=\cos(\omega+f)/\cos \phi,  \\
&\sin(\alpha-\Omega)=\sin(\omega+f)\cos i/\cos \phi,  \\
&\sin \phi= \sin i \sin(\omega +f), 
\end{aligned}
\right.
\end{align}
where $\alpha$ stands for the right ascension of the satellite (again, we refer to \cite{wKa66} for omitted details). 
The longitude $\lambda$ is written
as a function of $\alpha$ and the sidereal time $\theta$ as 
\begin{align}
\lambda = \alpha - \theta 
= 
(\alpha - \Omega) + (\Omega - \theta).
\end{align}
The sidereal time $\theta$ evolves linearly with time as $\theta=\varpi_{\Earth}t$, with 
$\varpi_{\Earth} = 2 \pi / \textrm{sidereal day}$.
Writing the inverse of the radius as
\begin{align}
\frac{1}{r}=\frac{1 + e \cos f}{a(1-e^{2})},
\end{align} 
the quantities $\sin f$ and $\cos f$ are then written using their infinite series representation as a function of the mean anomaly  $M$ and the Bessel functions $J_{s}$ (see, \eg \cite{cMu99})
\begin{align}\label{eq:cosfsinf}
\left\{
\begin{aligned}
&\sin f= \lim_{k \to +\infty}2\sqrt{1-e^{2}}\sum_{s=1}^{k}
\frac{1}{s} \frac{\dd }{\dd e}J_{s}(se) \sin sM,  \\
&\cos f= \lim_{k \to +\infty} -e +\frac{2(1-e^2)}{e}\sum_{s=1}^{k}J_{s}(se)\cos sM.
\end{aligned}
\right.
\end{align}
Applying the aforementioned substitutions into Eq.\,(\ref{eq:VTess}) transforms it into an expression dependent solely on the orbital elements $(a,e,i,\Omega,\omega,M)$ and the sidereal time $\theta$. 
The angles appear as linear combinations over the rationales of
$M,   \theta - \Omega$ and $\omega$ \cite{wKa66}.  
Computing at this stage the brute-force $M$-average to derive the secular tesseral contribution would suppress the dynamical effects of the resonant terms for $12$-hour orbits. In fact, 
in the vicinity of  $12$-hour orbits, 
the fast angle
\begin{align}
u_{\F} = \theta - \Omega,
\end{align}
combines with the fast variable $M$ 
as
\begin{align}
2  u_{\textrm{S}}=M - 2 u_{\F},
\end{align}
to form a slow varying quantity. Therefore, in the neighborhood of $12$-hour orbits, 
the variable $u_{\textrm{S}}$ needs to be  considered as a slow and independent variable. 
Dealing therefore with the variables $u_{\F}, u_{\textrm{S}}, \omega$, there is one fast angle $u_{\F}$ and two slow angles, $\omega$ and $u_{\textrm{S}}$. The resonant tesseral contribution is therefore obtained by averaging over the fast angle $u_{\F}$ as
\begin{align}
\bar{V}_{\mathcal{T}} = \frac{1}{2\pi} \int_{0}^{2\pi}
V_{\mathcal{T}} \, \dd u_{\textrm{F}}.
\end{align}  
The final expression has the form
\begin{align}
\bar{V}_{\mathcal{T}} = \sum_{k=(k_{1},k_{2}) \in K}
h_{k}(a,e,i) \cos(\sigma_{k} + k_{1}\lambda_{lm}), \, K \subset \mathbb{Z}^{2},
\end{align} 
where  
\begin{align}
\sigma_{k} = k_{1} u_{\textrm{S}} + k_{2} \omega,
\end{align}
and $\lambda_{lm}$ is a constant phase-term
defined as 
\begin{align}
\left\{
\begin{aligned}
&c_{l,m} = -J_{lm} \cos m \lambda_{lm},\\ 
&s_{l,m} = -J_{lm} \sin m \lambda_{lm}.
\end{aligned}
\right.
\end{align}

In Tab.\,\ref{tab:ResCoeff}, we provide the final formal expression of the secular resonant terms\footnote{Note the discrepancies with the formula presented in \cite{aCe14} for the coefficients of the trigonometric terms with arguments $M - 2\theta_{S}+ 2 \Omega - 2 \omega$, $M - 2\theta_{S}+ 2 \Omega - 3 \omega$ and $M - 2\theta_{S}+ 2 \Omega +4 \omega $. We kindly acknowledge the authors of \cite{aCe14} for their independent confirmation.}
for $12$-hour orbits up to $l=4$ appearing in $\bar{V}_{\mathcal{T}}$ for the uplets $k \in K$ with Eq.\,(\ref{eq:cosfsinf}) truncated to $k_{\max}=4$.\\

Now that we have at hand the secular disturbing functions,  the dynamics  is cast into  a Hamiltonian framework 
accounting for the Keplerian central part,
\begin{align}\label{eq:H}
\mathcal{H}=
\mathcal{H}_{\textrm{kep}.}
+ V_{J_{2}} 
+ \mathcal{T}_{2},
\end{align}
where  $\mathcal{T}_{2}$ is obtained from (\ref{eq:VTess}) by restricting the expansion to $l=2$. 
The Hamiltonian must be a function of canonical variables
which are presented hereafter. We mention however that we might sometimes refer to quantities expressed in orbital elements (non-canonical elements) and it is understood that the elements are themselves function of canonical variables.

\begin{table}
	\centering
	\begin{tabular}{lcc}
		\hline
		\hline
		$k=(k_{1},k_{2})$ & $h_{k}(a,e,i)$ & $\sigma_{k}$  \\
		\hline
		\hline
		& &  \\	
		$(2,0)$ & $\frac{\mu  r_{\Earth}^2 J_{22}  9e (9e^{2}+8)  \sin^2 i 
		}{32 a^3}$ & $ M  - 2\theta_{S}+ 2 \Omega$  \\
		& &  \\
		$(2,2)$ & $\frac{\mu  r_{\Earth}^2 J_{22}  3 e (e^2-8)  (\cos i+1)^2 
		}{64 a^3}$ & $M  - 2\theta_{S}+ 2 \Omega + 2 \omega$   \\	
		& &  \\
		$(2,-2)$ & $\frac{ \mu  r_{\Earth}^2 J_{22}  e^{3} (\cos i-1)^{2} 
		}{64 a^3}$& $M  - 2\theta_{S}+ 2 \Omega - 2 \omega$  \\
		& &  \\
		\hline
		& &  \\
		$(2,-3)$ & $\frac{-\mu  r_{\Earth}^3 J_{32}  5 e^{4}  (\cos i-1)^2 \sin i
		}{1024 a^4}$ & $M  - 2\theta_{S}+ 2 \Omega - 3 \omega$  \\
		& &  \\		
		$(2,-1)$ & $ \frac{\mu  r_{\Earth}^3 J_{32}  15 e^{2} (49e^{2}+22)  (3\cos i +1)  (\cos i-1) \sin i
		}{128 a^4}$ & $M  - 2\theta_{S}+ 2 \Omega -  \omega$ \\
		& &  \\		
		$(2,1)$ & $\frac{-\mu  r_{\Earth}^3 J_{32}  
			15(239e^{4}+128e^{2}+64)
			(3\cos i-1)  (\cos i+1) \sin i
		}{512 a^4}$ & $M  - 2\theta_{S}+ 2 \Omega + \omega$ \\
		& &  \\		
		$(2,3)$ & $\frac{ \mu  r_{\Earth}^3 J_{32}  
			5e^{2}(e^{2}+6)
			(\cos i+1)^2 \sin i
		}{128 a^4}$ & $M  - 2\theta_{S}+ 2 \Omega + 3  \omega$ \\
		& &  \\
		\hline
		& &  \\
		$(2,0)$ & $\frac{-\mu  r_{\Earth}^4 J_{42}  
			75e(27e^{2}+8)
			(21\cos^{2}i\sin^{2}i-7\sin^{2}i-4\cos^{2}i+4) 
		}{256 a^5}$ & $M  - 2\theta_{S}+ 2 \Omega $  \\
		& &  \\	
		$(2,-2)$ & $\frac{\mu  r_{\Earth}^4 J_{42}  
			245e^{3}(\cos i -1)
			(7\cos i\sin^{2}i-\cos i +1)}{128 a^5}$ & $M  - 2\theta_{S}+ 2 \Omega -2 \omega $  \\
		& &  \\	
		$(2,2)$ & $\frac{\mu  r_{\Earth}^4 J_{42}  
			15e(33e^{2}+8)(\cos i +1)
			(7\cos i\sin^{2}i-\cos i -1)}{128 a^5}$ & $M  - 2\theta_{S}+ 2 \Omega +2 \omega $  \\
		& &  \\	
		$(2,4)$ & $\frac{\mu  r_{\Earth}^4 J_{42}  
			35e^{3}(\cos i +1)^{2} \sin^{2}i}{512 a^5}$ & $M  - 2\theta_{S}+ 2 \Omega +4 \omega $  \\
		& &  \\	
		$(4,0)$ & $\frac{\mu  r_{\Earth}^4 J_{44}525e^{2}(31e^{2}+12)\sin^4 i}{32a^{5}}$ & $2(M  - 2\theta_{S}+ 2 \Omega) $  \\
		& &  \\	
		$(4,2)$ & $\frac{\mu  r_{\Earth}^4 J_{44}  
			105(65e^{4}+16e^{2}+16)(\cos i +1)^{2} \sin^{2} i}{64 a^5}$ & $2(M  - 2\theta_{S}+ 2 \Omega) +2 \omega $  \\
		& &  \\	
		$(4,4)$ & $\frac{-\mu  r_{\Earth}^4 J_{44}  
			35e^{2}(2e^{2}-3)(\cos i +1)^{4}}{32 a^5}$  & $2(M  - 2\theta_{S}+ 2 \Omega) +4 \omega $  \\
	\end{tabular}	
	\caption{\label{tab:ResCoeff}
		Formal coefficients and resonant angles
		of the $2:1$ resonance up to $l_{\max}=m_{\max}=4$
		and up to the $4$th order in eccentricity. 
	}
\end{table}

%======
\subsection{Dynamics}
%======
We start by introducing  the following canonical resonant coordinates  
\begin{align}\label{eq:ResCoord}
\left\{
\begin{aligned}
& I_{1}=-L, \hspace{1.56cm} u_{1}=2\theta-\ell -2h, \\
& I_{2}=G, \hspace{1.8cm} u_{2}=g,\\
& I_{3}=H-2L, \hspace{0.9cm} u_{3}=h, \\
& I_{4}=-2L-\Gamma, \hspace{0.72cm} u_{4}=-\theta,
\end{aligned}
\right.
\end{align}
where $(L,G,H,\ell,g,h)$ denote the classical canonical Delaunay variables related to the Keplerian elements
by 
\begin{align}
\left\{
\begin{aligned}
& L=\sqrt{\mu \, a}, \hspace{1.68cm} \ell=M, \\
& G=L \, \sqrt{1-e^{2}}, \hspace{0.77cm} g=\omega,\\
& H= G \, \cos i, \hspace{1.24cm} h=\Omega.
\end{aligned}
\right.
\end{align}
Given that the Hamiltonian (\ref{eq:H}) is time-dependent,
$
\dot{\theta}=\varpi_{\Earth}=2\pi/\textrm{sidereal day}$,
we  supplement the dynamics with $1$-\DOF given by the canonical conjugate variables
denoted $(\Gamma,\tau=\theta)$, with $\dot{\theta}=\dot{\tau}=\varpi_{\Earth}$. Those variables 
enter into the definition of $(I_{4},u_{4})$. The autonomous 
dynamics (we still note $\mathcal{H}$ the new Hamiltonian) reads
\begin{align}\label{eq:Haut}
\mathcal{H}=\mathcal{H} + \varpi_{\Earth}\Gamma.
\end{align}
The Hamiltonian (\ref{eq:Haut}) written in terms of the resonant coordinates (\ref{eq:ResCoord}) reduces  to a 2-\DOF system 
as  both $u_{3}$ and $u_{4}$ are ignorable. Consequently, their conjugate canonical actions, $I_{3}$ and $I_{4}$, are constant over time (\ie parameters). 
%For the $l=2$ terms of $H_{\mathcal{T}_{2}}$, the angles appearing in the 2-\DOF systems are given by $u_{1}$, $u_{1}+2u_{2}$ and $u_{1}-2u_{2}$.
Note that when $\dot{u_{2}}=0$, the problem is a $1$-\DOF problem and is therefore 
trivially
integrable. 

%==============
\subsubsection{The integrable approximation}
\begin{figure}
	\begin{center}	
		\includegraphics[width=0.6\linewidth]{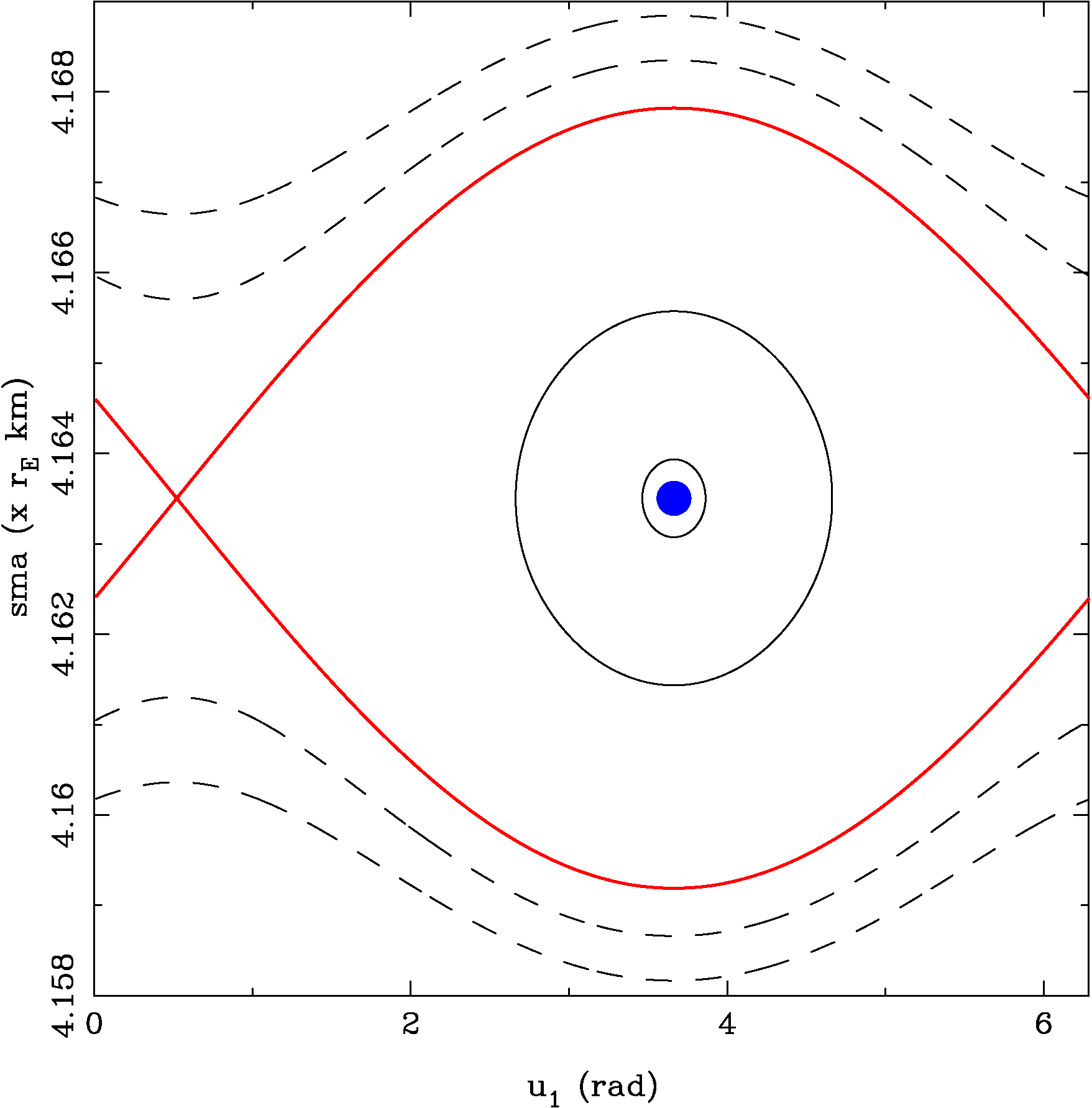}
	\end{center}		
	\caption{\label{fig:1dof}
		Phase space  of the resonant integrable approximation.  
		The width of the separatrix (red curve) allows 
		excursion of the semi-major axis up to $2\Delta a= 54$ km within the libration domain. The oscillations near the elliptic fixed-point (blue point) have a period of about $1.76$ years.
	}
\end{figure}

%==============
When  $\dot{u_{2}}\neq0$, we derive 
the resonant integrable approximation assuming that the resonances are isolated \cite{aMo02}. It amounts to take into  account in  (\ref{eq:H}), besides the action-only dependent part, the harmonic   with the largest amplitude. For  Molniya's orbital parameters, the numerical evaluations of the coefficients gives
\begin{align}
\left\{
\begin{aligned}
& \vert h_{2,0}\vert \sim 3 \, \vert h_{2,2}\vert, \notag \\
& \vert h_{2,0}\vert \sim 10^{3} \, \vert h_{2,-2}\vert.
\end{aligned}
\right.
\end{align}
The $1$-\DOF approximation is  therefore  built  on the one-harmonic Hamiltonian 
\begin{align}\label{eq:1dofharmonic}
\tilde{\mathcal{H}}=\mathcal{H}_{0}+h_{2,0} \cos(u_{1}+2\lambda_{22}).
\end{align}
The resonance $\dot{u_{1}} =0$ (let us recall $u_{1}=2\theta-\ell -2h$), that we denote $\mathcal{R}_{u_{1}}$, occurs for
\begin{align}
\varpi_{0}(I_{1})=\partial_{I_{1}} \mathcal{H}_{0}=0.  
\end{align} 
Solving this equation  in $I_{1}$ for   $I_{2}, I_{3}$ determined by $\oe_{\mathcal{M}}$, 
we find a resonant action $I_{1}^{\star}$ leading to the resonant semi-major axis
\begin{align}
a_{\star}(I_{1}^{\star})=26,555 \, \textrm{km}.
\end{align} 
The orbits of (\ref{eq:1dofharmonic})
coincide with the set of level curves. Yet, as it will be clear in the subsequent sections, 
the phase space is analogue to the classical pendulum dynamics. 
Analytical characteristics of the resonance  might be derived from a low-order Taylor expansion of the Hamiltonian near $I_{1}=I_{1}^{\star}$.  By keeping only the 
quadratic action term,
it reduces the  Hamiltonian to
\begin{align}\label{eq:H1dofTaylor}
\tilde{\mathcal{H}}=\frac{1}{2}\alpha_{0}J_{1}^{2} + h_{2,0} \cos(u_{1}+2\lambda_{22}),
\end{align}
where $J_{1}=I_{1}-I_{1}^{\star}$ and 
\begin{align}
\alpha_{0}=\partial^{2}_{I_{1}I_{1}}\mathcal{H}_{0}\vert_{I_{1}=I_{1}^{\star}}.
\end{align}
The equilibria  are given by the solution of
\begin{align}\label{eq:vEq} 
\left\{
\begin{aligned}
&\dot{J_{1}}=-\partial_{u_{1}}\tilde{\mathcal{H}}=h_{2,0}\sin(u_{1}+2\lambda_{22})=0, \\
&\dot{u_{1}}=\partial_{J_{1}}\tilde{\mathcal{H}}=\alpha_{0}J_{1}=0,
\end{aligned}
\right.
\end{align}
leading to two the equilibrium solutions
\begin{align}\label{eq:Eq} 
\left\{
\begin{aligned}
&\bold{x}_{s}=(0,u_{s}=-2\lambda_{22})\simeq (0,3.66), \\ &\bold{x}_{u}=(0,u_{s}=\pi-2\lambda_{22}) \simeq (0,0.52).
\end{aligned}
\right.
\end{align}
The eigensystem of the Jacobian matrix associated to (\ref{eq:vEq}) evaluated at the equilibrium solutions (\ref{eq:Eq})  shows 
that $\bold{x}_{s}$ is elliptic (stable fixed point) and  $\bold{x}_{u}$ is  a saddle (unstable fixed point), from which emanates the separatrix (curve associated to the energy level of the unstable equilibria).  It also provides further temporal characteristic timescales. 
The eigensystem of the Jacobian 	
evaluated at $\bold{x}_{s}$ provide the two complex conjugate eigenvalues $(\lambda_{s},\bar{\lambda}_{s})$ leading to
the characteristic 
periods of libration in the harmonic regime 
\begin{align}
T_{\textrm{lib.}}=\frac{2 \pi}{\vert \Im(\lambda_{s})\vert} \simeq 1.76 \, \textrm{years}.
\end{align}
The eigensystem of the Jacobian evaluated at $\bold{x}_{u}$ is composed by two real eigenvalues $(\lambda_{u},-\lambda_{u})$
defining the e-folding time 
\begin{align}
T_{e} = 1/\vert \lambda_{u} \vert\simeq 0.28 \, \textrm{year}.
\end{align}	
The resonance half-width $\Delta J_{1}$ associated to (\ref{eq:H1dofTaylor}), \ie the distance between $J_{1}=0$ and the apex of the separatrix  satisfies  
\begin{align}
\tilde{\mathcal{H}}(\Delta J_{1},u_{s})=
\tilde{\mathcal{H}}(0,u_{u}), 
\end{align}
that is
\begin{align}
\frac{1}{2}\alpha_{0}\Delta J_{1}^{2} +h_{2,0}=-h_{2,0}.
\end{align} 
Solving the last equality for $\Delta J_{1}$, we find
\begin{align}
\Delta J_{1}= 2 \sqrt{\frac{\vert h_{2,0}\vert }{\vert \alpha_{0} \vert }} \leftrightarrow \Delta a = 27 \, \textrm{km},
\end{align}
where the reader is referred to Fig.\,\ref{fig:1dof} for further qualitative details.

%==============
\subsubsection{The $2$-\DOF picture}\label{subsec:2dofGeo}
%==============
When $\dot{u_{2}} \neq 0$, 
the energy function (\ref{eq:H}) defines a
$2$-\DOF problem with 
the
multiplet of three resonances 
$\mathcal{R}_{u_{1}}, \mathcal{R}_{u_{1}+2g}, \mathcal{R}_{u_{1}-2g}$. Each isolated resonant problem admits its own pendulum reduction, with the possibility to overlap \cite{bCh79}.   
Analytical insights might be gained by some simplifications. In fact, let us approximate (\ref{eq:H}) with the following $2$-\DOF problem
\begin{align}\label{eq:K}
\mathcal{K} = 
\frac{1}{2}\alpha_{0}J_{1}^{2}
+ \varpi_{g} \Gamma 
+ h_{2,0}  \cos (\phi) 
+ h_{2,-2} \cos (\phi+2 \tau)
+ h_{2,2}  \cos (\phi-2 \tau),
\end{align} 
with $\phi=u_{1}+2\lambda_{22}$, where
we have assumed the rate of variation of $u_{2}=g$ to be ruled by the $\mathcal{H}_{J_{2}}$ part, that is 
\begin{align}\label{eq:FreqJ2omega}
\left.
\dot{u_{2}}\equiv\varpi_{g}
=\frac{\partial \mathcal{H}_{J_{2}}}{\partial G}=
\frac{3}{4}r_{\Earth}^{2}J_{2}\frac{\mu^{1/2}}{a^{7/2}}
\frac{5 \cos^{2}i -1}{(1-e^{2})^{2}}
\right|_{\oe_{\mathcal{M}}}.
\end{align}
Therefore, $u_{2}$   evolves linearly with time which we denote as
$\tau$. 
Using the canonical equations, we 
find the three resonances centers of
$
\mathcal{R}_{u_{1}}, 
\mathcal{R}_{u_{1}+2g}, 
\mathcal{R}_{u_{1}-2g}$
to be located 
respectively at 
\begin{align}\label{eq:center}
\left\{
\begin{aligned}
&c_{u_{1}}(J_{1})=0,  \\
&c_{u_{1}+2g}(J_{1})=-2\varpi_{g}/\alpha_{0},	\\
&c_{u_{1}-2g}(J_{1})=2\varpi_{g}/\alpha_{0}. 
\end{aligned}
\right. 	
\end{align}
The mutual distances of the center of the resonances with respect to the center of $\mathcal{R}_{u_{1}}$,
\begin{align}\label{eq:Distance}
\left\{
\begin{aligned}
&\delta(\mathcal{R}_{u_{1}},\mathcal{R}_{u_{1}+2g})=\vert c_{u_{1}} - c_{u_{1}+2g}\vert = 2 \vert \varpi_{g}/\alpha_{0} \vert ,  \\
&\delta(\mathcal{R}_{u_{1}},\mathcal{R}_{u_{1}-2g})=\vert c_{u_{1}} - c_{u_{1}-2g}\vert = 2 \vert \varpi_{g}/\alpha_{0}\vert ,
\end{aligned}
\right. 	
\end{align}
are  small 
given 
that $i_{\mathcal{M}} \sim i_{\star}$, 
where $5 \cos^{2} i_{\star} -1 =0$ (critical inclination value). The corresponding $\delta a$  amounts to be less than $1$ km.  
Treated as isolated, 
the resonances 
$
\mathcal{R}_{u_{1}}, 
\mathcal{R}_{u_{1}+2g}, 
\mathcal{R}_{u_{1}-2g}$
have the respective half-widths
\begin{align}
\left\{
\begin{aligned}
&\Delta_{\mathcal{R}_{u_{1}}}J_{1} = 2\sqrt{\vert h_{2,0}\vert/\vert \alpha_{0}\vert}
\leftrightarrow \Delta_{\mathcal{R}_{u_{1}}} a =
27.5 \,
\textrm{km},  \\
&\Delta_{\mathcal{R}_{u_{1}+2g}}J_{1}
=
2\sqrt{\vert h_{2,2}
	\vert/\vert \alpha_{0}\vert}
\leftrightarrow \Delta_{\mathcal{R}_{u_{1}+2g}} a
=14.4 \,
\textrm{km}
, \\
&\Delta_{\mathcal{R}_{u_{1}-2g}}J_{1} =2\sqrt{\vert h_{2,-2}
	\vert/\vert \alpha_{0}\vert}
\leftrightarrow \Delta_{\mathcal{R}_{u_{1}-2g}}a
= 0.78  \,
\textrm{km}
.
\end{aligned}
\right. 	
\end{align}  
As inferred from the numerical computation of $h_{2,-2}$, the resonance 
$\mathcal{R}_{u_{1}-2g}$ is negligible for 
practical purposes. Due to the inequalities
\begin{align}  
\left\{
\begin{aligned}
\Delta_{\mathcal{R}_{u_{1}}} +  \Delta_{\mathcal{R}_{u_{1}+2g}} \gg 
\delta(\mathcal{R}_{u_{1}},\mathcal{R}_{u_{1}+2g}), \\
\Delta_{\mathcal{R}_{u_{1}}} +  \Delta_{\mathcal{R}_{u_{1}-2g}} \gg
\delta(\mathcal{R}_{u_{1}},\mathcal{R}_{u_{1}-2g}),
\end{aligned}
\right.
\end{align}
a complete resonance overlap takes place (\ie the resonances are strongly overlapped), by which is meant that the widths of the resonances (treated as isolated) are much larger than their mutual separations. This paradigm is encapsulated into an analogue of the so-called \textit{modulated pendulum approximation} (see \eg \cite{aMo02}). From this analogy, we might infer the absence of large chaotic seas known to exist for similar eccentricity range but at lower inclination  \cite{jDa15,tEl96}. Instead, we expect chaotic motions to appear only in the vicinity of the unperturbed separatrix \cite{aMo02,nMu97}, with a librational region filled by stable orbits. This fact is indeed corroborated by computing the Poincar\'e map.\\

\begin{figure}
	\begin{center}
		\includegraphics[width=0.9\linewidth]{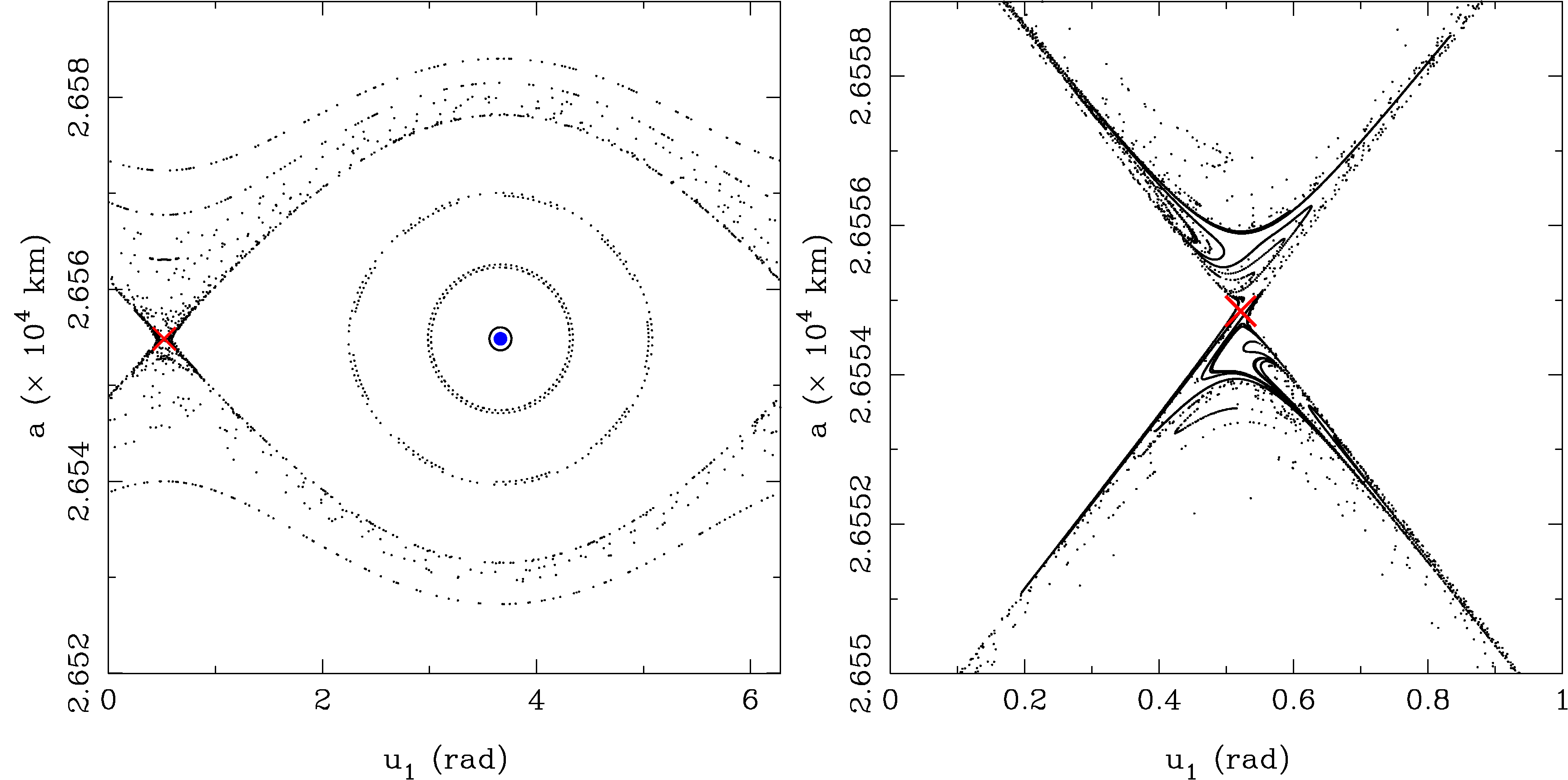}
		\caption{\label{fig:section}
			(Left) Poincar\'e section associated to (\ref{eq:K})
			computed for $g(0)=0$. The unstable fixed point is labeled with the red cross, the blue circle 
			surrounds the stable periodic orbit. 
			The phase space is similar to the integrable approximation but contains  a thin chaotic layer (scattered erratic points) surrounding the unperturbed separatrix.  Each considered initial condition has been iterated $100$ times under  $\mathcal{P}$. (Right) Details of finite pieces of the stable manifold $\mathcal{W}^{s}(\bold{x}_{u})$. 
		}
	\end{center}
\end{figure}

\noindent \textbf{Stroboscopic map.}
The Hamiltonian (\ref{eq:K}) is a 1-\DOF system periodically perturbed. Its phase space can be described by computing the associated Poincar\'e map, which is, given the periodic nature of the forcing, a stroboscopic mapping \cite{jMe92,sWi03}. Let us denote this mapping by $\mathcal{P}$ and by $\mathcal{V}(0)$ a neighborhood of $J_{1}=0$. 
By defining the lift and projector operators respectively  as  
\begin{align}
\frak{l}: \mathcal{V}(0) \times [0,2\pi] &\to \mathcal{V}(0) \times B \times [0,2\pi]^{2}, \, B \subset \mathbb{R},\notag \\
z=(J_{1},u_{1})  &\mapsto \frak{l}(z)=\bold{x}=(J_{1},\Gamma,u_{1},\tau),
\end{align}
and 
\begin{align}
\frak{p}:  \mathcal{V}(0) \times B \times [0,2\pi]^{2} & \to  \mathcal{V}(0) \times [0,2\pi], \notag \\
\bold{x}=(J_{1},\Gamma,u_{1},\tau)  & \mapsto \frak{p}(\bold{x})=(J_{1},u_{1}),
\end{align}
the stroboscopic map is defined as  
\begin{align}
\mathcal{P}: \mathcal{V}(0) \times [0,2\pi] & \to  \mathcal{V}(0) \times [0,2\pi], \notag \\
z & \mapsto \mathcal{P}(z)=z'=
\frak{p}\circ \Phi^{T_{g}} \circ \frak{l} 
(z),
\end{align}
where $\Phi^{t}$ is the flow at time $t$ associated to (\ref{eq:K}) and 
$
T_{g} = 2\pi/\varpi_{g}.
$
Note that the lift is parameterised by the choice of $\tau(0)=g_{0}$. The  ``dummy'' variable $\Gamma$ does not enter into the equations of motion. 
For Molniya-like spacecraft, $T_{g}$ defines a period of about $100$ years (\ie the order of  $ 10^{4}$ orbital revolutions). 
The mapping $\mathcal{P}$ is constructed numerically based on the numerical propagation of the system (\ref{eq:K}). 
Given a value of $g_{0}$, the 
coordinates of the fixed points of the mapping $\mathcal{P}$ (\ie the periodic orbits of (\ref{eq:K})) are  determined using a Newton method.   
Due to the periodicity 
\begin{align}
\mathcal{K}(J_{1},u_{1};\tau) = \mathcal{K}(J_{1},u_{1};\tau+\pi),
\end{align}
the domain of $g$ can be restricted to $[0,\pi]$. 
Let us recall that a fixed point $z_{\star}$ of $\mathcal{P}$, $\mathcal{P}(z_{\star})=z_{\star}$, is hyperbolic when the linearisation has at least one eigenvalue with modulus greater than one. In case of complex eigenvalues, the fixed point is elliptic. For $g_{0}=0$, the two fixed points (semi-major axis given in km)  read as 
\begin{align}
\left\{
\begin{aligned}
&\bold{x}_{s}=(a,u_{1})=(26554.841,3.662), \\
&\bold{x}_{u}=(a,u_{1})=(26554.850,0.521).
\end{aligned}
\right.
\end{align} 
Changing $g_{0}$ alters slightly those coordinates and the slope of the eigenvectors associated to the unstable periodic orbit, which may widen the aperture 
of the librational domain
by a few kilometers.
The stable and unstable manifolds associated to an hyperbolic point $z_{\star}$, 
\begin{align}
\left\{
\begin{aligned}
&\mathcal{W}^{s}(z_{\star})=\{z \, , \, \norm{\Phi^{t}(z)-z_{\star}} \to 0, \, t \to +\infty\}, \notag \\
&\mathcal{W}^{u}(z_{\star})=\{z \, , \, \norm{\Phi^{-t}(z)-z_{\star}} \to 0, \, t \to +\infty\},
\end{aligned}
\right.
\end{align}
are grown by iterating points belonging to the fundamental domain $I \subset E^{s,u}$, where $E^{s,u}$ are respectively the stable and unstable eigenspaces  associated to $z_{\star}$ (and derived from the eigensystem analysis). Recall that 
$\mathcal{W}^{s,u}$ are locally tangent to $E^{s,u}$.
In Fig.\,\ref{fig:section}, we show the Poincar\'e section containing a chaotic zone surrounding the ``unperturbed separatrix''. A smaller portion of the phase space  shows the first lobes associated to the stable manifold.    
The analysis of the eigensystem associated to the linearisation of  $\mathcal{P}$ at the saddle fixed-point is 
enlightening in deriving the Lyapunov timescale analytically.  
Let us 
recall that a Floquet characteristic exponent $\mu$ is a complex number satisfying 
\begin{align}
\lambda = e^{\mu T_{g}},
\end{align}
where $\lambda$ is an eigenvalue associated to the linearisation $\textrm{D}\mathcal{P}$ about 
the fixed point.
For the hyperbolic saddle, the two eigenvalues 
$\{\lambda_{1},\lambda_{2}=1/\lambda_{1}\}$ are real
and so are the corresponding $\{\mu_{1},\mu_{2}\}$, called in this case the Lyapunov exponents. 
From
\begin{align}
\mu = T_{g}^{-1} \log \lambda, 
\end{align}
the timescale of $1/\mu \sim 17$ years
is derived for the largest  eigenvalue. This timescale has been compared with a brute-force estimation of the maximal Lyapunov exponents $\chi$ based on the variational dynamics (and their  associated Lyapunov time $\tau_{\mathcal{L}}=1/\chi$)  in the vicinity of the hyperbolic saddle. For hyperbolic orbits, we found Lyapunov times in the range of $15-18$ years, thus in very good agreement with the analytical  timescale based on D$\mathcal{P}$. \\

\noindent \textbf{Consequences of the chaotic layer.}
The presence of  the thin chaotic layer surrounding the unperturbed separatrix brings  important distinguishable qualitative features to the dynamics: the semi-major axis might display 
intermittency phenomena and the resonant angle alternates between librational and circulational regimes.
We note that such features have been observed for 
simulated geosynchronous orbits  \cite{sBr05,iWy07}. More precisely, for initial conditions in the chaotic layer, the orbit ``swaps'' between the ``inner-libration''  regime, characterised by
\begin{align}
\langle J_{1}\rangle_{u_{1}}^{\textrm{lib.}} \simeq 0,
\end{align}
and the ``outer-circulation'' regime for which 
\begin{align}
\langle J_{1}\rangle_{u_{1}}^{\textrm{circ.}} \not\simeq 0.
\end{align} 
The alternation takes place when the orbit returns close enough to the hyperbolic saddle $\bold{x}_{u}$  where the scattering takes place.

\subsubsection*{Remark}
\begin{figure}
	\centering
	\includegraphics[width=0.9\textwidth]{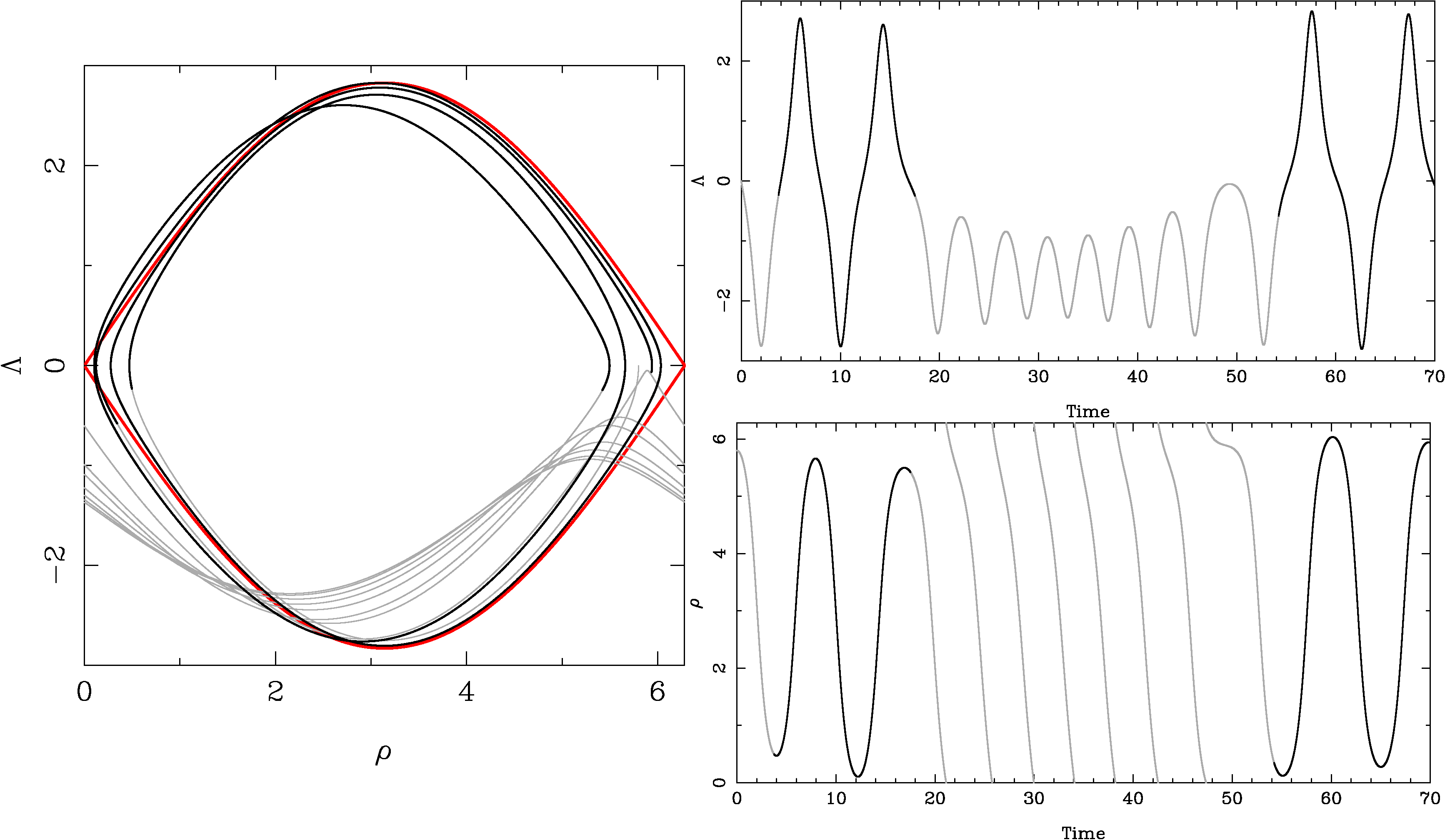}
	\caption{\label{fig:sketch}
		Composite plot illustrating the mechanisms of the intermittency phenomena. The 
		red line represents the separatrix of the integrable model $\mathcal{T}$. Realisations of the stable and unstable manifolds,
		for $\epsilon \neq 0, \, \epsilon \ll 1$,
		are not shown for the sake of readability. One hyperbolic orbit trapped in the hyperbolic tangle is highlighted in the phase space, with a color code depending on the regime of the resonant angle. When the resonant angle circulates 
		(grey color),  the  action 
		takes negative $\Lambda$'s.
		When the angle librates 
		(black color),  the action variable performs the full homoclinic loop and exhibit larger variations.	
	}
\end{figure}

	This mechanism is illustrated and summarised within the composite panel in Fig.\,\ref{fig:sketch} based on the Hamiltonian model 
	\begin{align}
	\mathcal{T}=\frac{1}{2}\Lambda^{2}
	+ \Lambda_{1} + \cos(\rho) + \cos(\rho+ \epsilon \rho_{1}), \epsilon \ll 1.
	\end{align}
	For $\epsilon=0$, $\mathcal{T}$ is integrable and has a saddle structure at  $(\Lambda,\rho)=(0,\pi)$.  The separatrix has a cat-eye topology with half-width 
	$\Delta = 2 \sqrt{2}$.
	When $\epsilon \neq 0, \epsilon \ll 1$, the 
	resonances 
	$\mathcal{R}_{\rho}$ and 
	$\mathcal{R}_{\rho+ \epsilon \rho_{1}}$,  
	$\epsilon$-apart, produce a  
	separatrix splitting. 
	The resonant angle of an orbit with initial condition in the hyperbolic set alternates among libration, 
	$\langle \Lambda \rangle_{\rho} \simeq 0$, 
	and circulation, 
	$\langle \Lambda \rangle_{\rho} \not\simeq 0$. The ``projection'' of one orbit with initial condition close to the saddle (trapped in the hyperbolic tangle) in the  space $(\Lambda,\rho)$ shows that the orbit remains mainly guided by the unperturbed separatrix. 
	Under our selected initial condition, when the angle circulates, the action
	is trapped in the tangle, evolving here in the domain 
	$\mathbb{\Lambda}^{-}:=\{\Lambda, \, \Lambda<0\}$.  When the angle librates, the action experiences  full homoclinic loops and evolve within 
	$
	\mathbb{\Lambda}=\mathbb{\Lambda}^{-}
	\cup
	\mathbb{\Lambda}^{+}$.  
	This process continues and possibly alternates in the vicinity of the  saddle, producing scattering and contributing to the growth of the tangent vector.\\

\noindent \textbf{Limitations of the model $\mathcal{K}$.}
The model (\ref{eq:K}) is based on geopotential perturbations only. 
To build a more realistic model, the lunisolar perturbations, Moon and Sun, need to be included. In its present form, model $\mathcal{K}$ is limited in two ways:
\begin{enumerate}
	\item Under the lunisolar effects and due to the proximity to the critical inclination value,  the hypothesis that the argument of the perigee $(g=u_{2})$  flows linearly with time at a (constant) rate given by the $J_{2}$ effect is violated (a fact also observed at the data level).   
	\item The assumption that both the eccentricity and inclination are parameters is no longer true under the influence of the lunisolar perturbation.
\end{enumerate}
Increasing the complexity of the model gradually, we overcome the first limitation by decoupling the equations of motions. We isolate a simplified energy function $\mathcal{L}$ that dictates the time evolution of the argument of perigee, 
$\dot{g}=\partial_{G}\mathcal{L}$, that we   use to form a $6$-dimensional dynamical system with constant eccentricity and inclination. The variables $(J_{1},u_{1})$  are then  studied.  The second limitation is raised by introducing  a $3$-\DOF Hamiltonian system, where both the eccentricity and inclination vary according to the dynamics.

\section{Secular Hamiltonian including lunisolar effects}\label{sec:LS}

\subsection{The doubly-averaged lunisolar Hamiltonian} 
We adopt a simplified  sub-model of the quadrupolar doubly-averaged formulation to model the external third-bodies perturbations. The quadrupolar  approximation is commonly employed to study medium-Earth orbit dynamics and has already demonstrated its relevance (see \eg \cite{iGk16,jDa18}). Starting from the Hamiltonians  
\begin{align}
\left\{
\begin{aligned}
&\mathcal{H}_{\Moon} = 
-\frac{\mu_{\Moon}}{r_{\Moon}}
\Big(
\frac{r_{\Moon}}{\norm{\bold{r}-\bold{r}_{\Moon}}}
-
\frac{\bold{r}\cdot \bold{r}_{\Moon}}{r_{\Moon}^{2}}
\Big), \notag \\	
&\mathcal{H}_{\Sun} = 
-\frac{\mu_{\Sun}}{r_{\Sun}}
\Big(
\frac{r_{\Sun}}{\norm{\bold{r}-\bold{r}_{\Sun}}}
-
\frac{\bold{r}\cdot \bold{r}_{\Sun}}{r_{\Sun}^{2}}
\Big),
\end{aligned}
\right.
\end{align}
where $\bold{r}_{\Moon},\bold{r}_{\Sun}$ denote the geocentric vectors of the Moon and the Sun, $r_{\Moon},r_{\Sun}$ the corresponding geocentric distances and $\mu_{\Moon},\mu_{\Sun}$ their respective gravitational parameters, a Legendre-like expansion of 
$\mathcal{H}_{\Moon}$ and 
$\mathcal{H}_{\Sun}$, truncated to $l=2$ (\textit{quadrupolar hypothesis}), and averaged over 
the mean anomalies $(M,M_{\Moon})$ and $(M,M_{\Sun})$ respectively, defines the so-called doubly averaged third-body model. 
This averaging is performed in closed form over the eccentricity. Contrarily to the inner-perturbative part (geopotential), it requires to use the differential relationship 
\begin{align}
\dd M = (1 - e \cos E) \,\dd E,
\end{align}    
coming from Kepler's equation ($E$ refers to the eccentric anomaly).
The double-averaging 
\begin{align}
\left\{
\begin{aligned}
&\mathcal{H}_{\Moon} = \frac{1}{(2\pi)^{2}}
\int_{0}^{2\pi}
\int_{0}^{2\pi}
\mathcal{H}_{\Moon} \, \dd M \, \dd M_{\Moon}
=
\frac{1}{(2\pi)^{2}}
\int_{0}^{2\pi}
\int_{0}^{2\pi}
\mathcal{H}_{\Moon} (1 - e \cos E) 
\frac{r_{\Moon}^{2}}{a_{\Moon}^{2}\sqrt{1-e_{\Moon}^{2}}}\,
\dd E \,
\dd f_{\Moon}, \notag \\	
&\mathcal{H}_{\Sun} = \frac{1}{(2\pi)^{2}}
\int_{0}^{2\pi}
\int_{0}^{2\pi}
\mathcal{H}_{\Sun} \, \dd M \, \dd M_{\Sun}
=
\frac{1}{(2\pi)^{2}}
\int_{0}^{2\pi}
\int_{0}^{2\pi}
\mathcal{H}_{\Sun} (1 - e \cos E) 
\frac{r_{\Sun}^{2}}{a_{\Sun}^{2}\sqrt{1-e_{\Sun}^{2}}}\,
\dd E \,
\dd f_{\Sun}, \notag 
\end{aligned}
\right.
\end{align}

reduces $\mathcal{H}_{\Moon}$ and $\mathcal{H}_{\Sun}$ to an expansion of the form
\begin{align}
\left\{
\begin{aligned}
&\mathcal{H}_{\Moon} = h_{0}^{\Moon}(a,e,i) + \sum_{j} h_{j}^{\Moon}(a,e,i) \cos \phi^{\Moon}_{j}, \notag \\	
&\mathcal{H}_{\Sun} =  h_{0}^{\Sun}(a,e,i) +
\sum_{j} h_{j}^{\Sun}(a,e,i) \cos \phi^{\Sun}_{j},
\end{aligned}
\right.
\end{align}
where $\phi^{\Moon}_{i}$ and $\phi^{\Sun}_{j}$ are permitted linear combinations of
$(\omega,\Omega,\Omega_{\Moon})$ and $(\omega,\Omega)$ respectively. Given that  the angle $\Omega_{\Moon}$ does not enter  $\phi^{\Sun}_{j}$, the summations  are homogenised  by introducing $\phi_{q}$, where 
\begin{align}
\phi_{q} = q_{1} \omega + q_{2} \Omega + q_{3} \Omega_{\Moon},
\end{align}
with the convention that $q_{3}=0$ for the permissible  solar arguments.
The quadrupolar doubly-averaged lunisolar Hamiltonian reads therefore 
\begin{align}\label{eq:HMS}
\mathcal{H}_{\Moon\Sun}
=
\mathcal{H}_{\Moon} + \mathcal{H}_{\Sun}
=
(h_{0}^{\Moon}+h_{0}^{\Sun}) + 
\sum_{q \in \mathcal{Q}} (
h_{q}^{\Moon}+h_{q}^{\Sun}
) \cos \phi_{q},
\end{align}
with (under the quadrupolar assumption)
\begin{align}
\mathcal{Q}=
\{
q \in \mathbb{Z}^{3}_{\star} \, \vert \, 
q_{1} \in \{-2,0,2\}, (q_{2},q_{3}) 
\in
\{-2,-1,0,1,2\}^{2}
\}.
\end{align}
For the sake of concision, let us denote $h_{j}^{\Moon\Sun}=h_{j}^{\Moon}+h_{j}^{\Sun}$.
The Hamiltonian (\ref{eq:HMS}) is in general non-autonomous as time enters through the ecliptic precession of the lunar node, well-approximated by the linear law (Moon's elements are referred to the ecliptic plane)
\begin{align}
\Omega_{\Moon} \simeq \Omega_{\Moon}(0) + \varpi_{\Omega_{\Moon}} t,
\end{align} 
where $2\pi/\vert \varpi_{\Omega_{\Moon}}\vert$ defines a period of about $18.6$ years. The Moon's inclination to the ecliptic plane is set to $i_{\Moon}=5^{\circ}15$. However, as we will see hereafter, the simplified model allowed by Molniya's parameters leads to a model independent of the argument of the Moon and, therefore, to an autonomous model. 

\subsection{Model for the time evolution of $\omega$}
The so-called ``double resonance model'' employed in \cite{tlZh15} based on the lunisolar harmonics $\cos 2g$ and $\cos 2g\pm h$ have shown to provide a realistic model to capture the time evolution of the argument of perigee. They compared orbits generated using this model against TLEs data on several cases and were able to reproduce qualitatively the time evolution of the argument of perigee on several decades. 
More recently,  \cite{tTa20} advocated  that the $\cos h$ term produces a significant contribution to the dynamics of $\omega$,
where the reader is referred to  Appendix \ref{app:ForceModel} for further details. 
We  adopt the following $2$-\DOF Hamiltonian system
\begin{align}\label{eq:L0}
\mathcal{L}(G,H,g,h)=
\mathcal{L}_{0}(G,H) +
\mathcal{L}_{1}(G,H,g,h),
\end{align}	
where
\begin{align}
\left\{
\begin{aligned}
&\mathcal{L}_{0}=\mathcal{H}_{J_{2}} 
+ h_{0}^{\Moon\Sun}, \notag \\
& \mathcal{L}_{1}=
h_{2g}^{\Moon\Sun} \cos(2g) 
+  h_{2g+h}^{\Moon\Sun} \cos(2g+h)
+  h_{2g-h}^{\Moon\Sun} \cos(2g-h),
\end{aligned}
\right.
\end{align}
to model the time evolution of $\omega$. From the canonical equations derived from (\ref{eq:L0}), we derive the dynamics of $g$.  The formal coefficients appearing in (\ref{eq:L0}), expressed using the Keplerian elements,  are listed in 
table \ref{tab:LunarSolar}. The terms $h_{0}^{\Moon}$ and $h_{0}^{\Sun}$ refer to the 
action dependent only terms of $\mathcal{H}_{\Moon}$ and $\mathcal{H}_{\Sun}$ and read
\begin{align}
\left\{
\begin{aligned}
&h_{0}^{\Moon}=\frac{\mu_{\Moon}a^{2}(3e^{2}+2)
	(3\sin^{2}i-2)(3\sin^{2}\varepsilon-2)(3\sin^{2}i_{\Moon}-2)}{64a_{\Moon}^{3}\eta_{\Moon}^{3}}\\ \notag
&h_{0}^{\Sun}=-\frac{\mu_{\Sun}a^{2}(3e^{2}+2)
	(3\sin^{2}i-2)(3\sin^{2}\varepsilon-2)}{32a_{\Sun}^{3}\eta_{\Sun}^{3}}.
\end{aligned}
\right.
\end{align}

\begin{table}
	\begin{minipage}{.45\textwidth}
		\begin{tabular}{cc}
			\hline
			\hline
			$h_{\sigma}^{\Moon}(a,e,i)$ & $\sigma$ \\
			\hline
			\hline
			& \\
			$-\frac{15 \mu_{\Moon}a^{2}e^{2}\sin^{2}i (3 \sin^{2}\varepsilon-2)(3\sin^{2}i_{\Moon}-2)}{64 a_{\Moon}^{3}\eta_{\Moon}^{3}}$ & $2g$   \\
			& \\
			$-\frac{15 \mu_{\Moon} a^{2}e^{2}(\cos i +1) \cos \varepsilon \sin i \sin \varepsilon(3 \sin^{2} i_{\Moon}-2)}{32 a_{\Moon}^{3}\eta_{\Moon}^{3}}$ & $2g+h$ \\
			& \\
			$-\frac{15 \mu_{\Moon} a^{2}e^{2}(\cos i -1) \cos \varepsilon \sin i \sin \varepsilon(3 \sin^{2} i_{\Moon}-2)}{32 a_{\Moon}^{3}\eta_{\Moon}^{3}}$ & $2g-h$ \\
			& \\
			$\frac{3 \mu_{\Moon} a^{2}(3e^{2}+2)
				\cos i \cos \varepsilon
				\sin i \sin \varepsilon (3 \sin^{2} i_{\Moon}-2)}{16 a_{\Moon}^{3}\eta_{\Moon}^{3}}$ & $h$
		\end{tabular}
	\end{minipage}
	\hspace{1cm}
	\begin{minipage}{.45\textwidth}
		\begin{tabular}{cc}
			\hline
			\hline
			$h_{\sigma}^{\Sun}(a,e,i)$ & $\sigma$ \\
			\hline
			\hline
			& \\
			$\frac{15 \mu_{\Sun}a^{2}e^{2}\sin^{2}i (3 \sin^{2}\varepsilon-2)}{32 a_{\Sun}^{3}\eta_{\Sun}^{3}}$ & $2g$   \\
			& \\
			$\frac{15 \mu_{\Sun}a^{2}e^{2}(\cos i +1) \cos \varepsilon \sin i \sin \varepsilon}{16 a_{\Sun}^{3}\eta_{\Sun}^{3}}$ & $2g+h$ \\
			& \\
			$\frac{15 \mu_{\Sun}a^{2}e^{2}(\cos i -1) \cos \varepsilon \sin i \sin \varepsilon}{16 a_{\Sun}^{3}\eta_{\Sun}^{3}}$ & $2g-h$ \\
			& \\
			$\frac{-3 \mu_{\Sun} a^{2}(3e^{2}+2)
				\cos i \cos \varepsilon
				\sin i \sin \varepsilon}{8 a_{\Sun}^{3}\eta_{\Sun}^{3}}$ & $h$
		\end{tabular}
	\end{minipage}
	\caption{\label{tab:LunarSolar}Formal expression of the lunar and solar coefficients associated  to the harmonics $2g$, $2g \pm h$ and $h$. The obliquity of the ecliptic with respect to the equatorial plane is $\varepsilon=23^{\circ}44$. The quantity $i_{\Moon}$ refers to the inclination of the Moon with respect to the ecliptic plane, $i_{\Moon}=5^{\circ}15$.}
\end{table}

\subsection{Effects of lunisolar perturbation on the tesseral dynamics} 
We investigate now how the lunisolar perturbation affects the hyperbolic structures of the tesseral problem for Molniya parameters.

The basic model dictating the time evolution of $g$ being established by (\ref{eq:L0}), we focus now on two dynamical systems improving the caveats of  (\ref{eq:K}): 
\begin{enumerate}
	\item First, we introduce the differential system in $\mathbb{R}^{6}$ defined by the equations of motion (EoM):
	\begin{align}\label{eq:modelJ}
	\left\{
	\begin{aligned}
	&\dot{J}_{1}=-\partial_{u_{1}}\mathcal{J}(J_{1},u_{1},g), \\
	&\dot{u}_{1}=\partial_{J_{1}}\mathcal{J}(J_{1},u_{1},g), \\
	&\dot{G}=-\partial_{g}\mathcal{L}(G,H,g,h), \\
	&\dot{g}=\partial_{G}\mathcal{L}(G,H,g,h),\\
	&\dot{H}=-\partial_{h}\mathcal{L}(G,H,g,h),\\
	&\dot{h}=\partial_{H}\mathcal{L}(G,H,g,h).
	\end{aligned}
	\right.
	\end{align}	
	where $\mathcal{J}$ is defined as,  
	\begin{align}
	\mathcal{J}(J_{1},u_{1},g(t))=\frac{1}{2}\alpha_{0}J_{1}^{2} + \mathcal{T}_{2}(u_{1},g(t)).
	\end{align}
	Here $\mathcal{L}$ is the basic lunisolar Hamiltonian function. For short, we refer to the EoM (\ref{eq:modelJ}) as  model $\mathcal{J}$
	and we denote the right-hand side by $v_{\mathcal{J}}$.
	\item Second, we consider the  $3$-\DOF Hamiltonian 
	\begin{align}\label{eq:S}
	\mathcal{S}(I_{1},I_{2},I_{3},
	u_{1},u_{2},u_{3})=\mathcal{H}_{\textrm{kep.}}(L)  + \varpi_{\Earth} \Gamma + \mathcal{T}_{2}(L,G,H,u_{1},g) + \mathcal{L}(G,H,g,h),
	\end{align}
	expressed within the resonant coordinates. With respect to the model $\mathcal{J}$, model $\mathcal{S}$  allows the action-terms to evolve under the correct dynamics. In particular, the tesseral coefficients $h_{2,0}(a,e,i)$, $h_{2,\pm2}(a,e,i)$ appearing in the dynamics of
	$(I_{1},u_{1})$
	are no longer frozen (instead, they  vary  according to the changes of the Delaunay action vector $(L,G,H)$). The right-hand side derived from $\mathcal{S}$ is denoted $v_{\mathcal{S}}$.
\end{enumerate}
Let us emphasise that both models are $\pi$-periodic in $g$. Molniya spacecraft have, in general, $g\sim 270^{\circ} \pm 20^{\circ}$.
For both models, in order to reveal the dynamical template
on
the $(I_{1},u_{1})$-plane,  we compute the Fast Lyapunov Indicators  \cite{mGu14,eLe16}  
on a 
$500 \times 500$ Cartesian mesh of initial conditions.  
We use the following definition of the FLI. 
For an $n$-dimensional autonomous ordinary differential system defined on a open domain $D \subset \mathbb{R}^{n}$,
$\dot{x}=f(x)$,  the FLI at time $t$ is derived 
from the linear map $\textrm{D}_{x}f$ at a point $x$:
\begin{align}
\textrm{D}_{x}f: \, &\mathbb{R}^{n} \to \mathbb{R}^{n}, \notag \\
&w \mapsto \textrm{D}_{x}f(x)w,  
\end{align}
and the associated  variational equations 
\begin{align}
\left\{
\begin{aligned}
& \dot{x}=f(x), \\
& \dot{w}= \textrm{D}_{x}f(x)w,
\end{aligned}
\right.
\end{align} 
as 
\begin{align}\label{eq:FLI}
\textrm{FLI}(t) = \sup_{0 \le \tau \le t} \log(\norm{w(\tau)}).
\end{align}
The vector $w \in \mathbb{R}^{n}$ denotes the tangent
(or deviation) vector. The computation of the FLIs over resolved grid of initial conditions discriminates efficiently the structures of a given dynamical system, including the stable or unstable manifolds, ordered or  chaotic seas. One advantage of the FLI over the characteristic Lyapunov exponent 
\begin{align}
\lambda(x,w) = \lim_{t \to +\infty} 
\frac{1}{t} \log(\norm{w(t)}),
\end{align}
is to get rid of the time-average computation, thus speeding the stability determination process.   
For regular orbits, the deviation vector grows linearly with time and  therefore the FLI on regular KAM tori are characterised by values close to $\log(\tau_{f})$.  In hyperbolic regions, the norm of the tangent vector grows exponentially fast, and therefore the FLI display a linear trend surpassing quickly the value taken on KAM objects (see \cite{dBe05}, chapter $5$, for perturbative estimates).

\subsubsection*{Remark}
	The parametric dependence 
	on $t$ in (\ref{eq:FLI}) is raised after  
	a calibration procedure. In our case, integration of several single orbits showed that $\tau_{f}=20$ years are sufficient to obtain a sharp distinction. 
	With our choice of units, the FLI  of regular orbits is characterised by the value 
	$\log(\tau_{f})=4.99$.

\subsubsection*{Remark}
	We restricted the  computations of the FLIs over $\Sigma$
	forward in time, \ie  on a time interval $[0,\tau_{f}]$, $\tau_{f}>0$, to obtain ``positive in time FLIs'',  
	$\textrm{FLIs}^{+}$. It is therefore understood that, in the context of the existence of hyperbolic invariants, these computations on $\Sigma$ would reveal the trace of the stable manifolds. 
	Similarly, backwards in time FLIs computed  over $[-\tau_{f},0]$, $\textrm{FLIs}^{-}$, 
	would reveal the trace of the unstable 
	manifolds.   Both manifolds  can be displayed on $\Sigma$ by plotting, \eg the standard average
	\begin{align}
	\textrm{FLI}=\frac{1}{2}(
	\textrm{FLI}^{+}+\textrm{FLI}^{-}),
	\end{align}
	or any others weighted average (see \eg \cite{mGu14,eLe16}). We computed a few of those maps backwards in time, to display the averaged FLI. However, because we are not particularly interested of highlighting  homoclinic connections,  we present hereafter only the forward in time FLI maps (\ie we display $\textrm{FLI}^{+}$).\\

The FLIs computation are performed for the  vector fields $v_{\mathcal{J}}$  and $v_{\mathcal{S}}$ 
over a Cartesian discretisation of $\Sigma \subset \mathbb{R}^{2}$, where
\begin{align}
\Sigma_{i_{0}} =
\big\{
(I_{1},I_{2},I_{3},u_{1},u_{2},u_{3}):
(I_{1},u_{1}) \in D,
u_{2}=270^{\circ},
u_{3}=0,  
I_{2}=0.7,
I_{3}(i_{0})=H_{\star}-2L_{\star}
\big\},
\end{align}
with 
\begin{align}
D = \mathcal{V}(I_{1}^{\star}) \times T, \quad T \subset [0,2\pi],
\end{align} 
and an initial deviation vector $w_{0}$ chosen arbitrarily\footnote{Instead of choosing a random vector $w_{0}$, we could have computed the FLIs over a basis of the tangent space. However, to avoid spurious structures, this refinement is not pursued herein.}.
The neighborhood of the resonant action $\mathcal{V}(I_{1}^{\star})$ represents typically a range of $70$ km in the semi-major axis.
Our numerical campaign is parametric through $4$ allowed  values of $i_{0}$, namely 
\begin{align}
i_{0}\in\{62.5^{\circ}, 63.4^{\circ}, 64.3^{\circ},65.2^{\circ}\},
\end{align}
``piercing'' the critical inclination value. This choice enters $\Sigma_{i_{0}}$ through 
\begin{align}
I_{3}(i_{0}) = L_{\star} \sqrt{1-e^{2}}
\cos i_{0} - 2 L_{\star}, \quad e=0.7.
\end{align}
Although aware that the precise geometry of the hyperbolic structures depend on the initial phasing $(\omega,\Omega)$, our investigations focus on  $(\omega,\Omega)=(270^{\circ},0)$. 
All the resulting maps of this numerical survey for models $\mathcal{J}$ and $\mathcal{S}$ are reported in Appendix \ref{app:maps} to ease the readability.  We show hereafter in composite panels only the relevant information for the analysis. From this survey, we observe that:
\begin{enumerate}
	\item Both models display a  saddle-like point in $\Sigma$.  This suggests the existence of an unstable periodic orbit (although this invariant has not been computed) for both flows, similar to the unstable periodic orbit we computed for model $\mathcal{K}$.
	Following this idea, the hyperbolic set (high values of FLIs with yellow color) emerging from the saddle-type structure is very likely to represent the intersections of  the stable manifold of the hyperbolic invariant with $\Sigma$. The fine mesh of initial conditions allows to recognise lobes distinctively for model $\mathcal{S}$.  
	\item The  model $\mathcal{J}$ is overall weakly perturbed, and the hyperbolic layer is very close to the unperturbed separatrix.  
	\item On the contrary, the hyperbolic layer of  $\mathcal{S}$ is much more developed.  
	This fact  is imputable to the indirect modulation of the coefficients $h_{2,0}(a,e,i)$ and $h_{2,\pm2}(a,e,i)$ under the lunisolar effects. We therefore see the signature of the lunisolar coupling onto the tesseral problem.    
	\item In general, the width of the hyperbolic layer  along a given line of $u_{1}$ increases with the values of $i_{0}$. For $u_{1}$ within the stable  librational domain (say $u_{1} \sim 3.6$), the 
	hyperbolic width along the line is about $1$ km large in the semi-major axis  ($i_{0}=62.5^{\circ}$) up to $\sim 6$ km large when $i_{0}=65.2^{\circ}$. For $u_{1}$ near the saddle, for the same inclination values, the hyperbolic layer foliates a width of about $10$ km in the semi-major axis for $i_{0}=62.5^{\circ}$ and up to $30$ km when $i_{0}=65.2^{\circ}$.
	\item Lastly, and more will be commented on that in the following, we notice a growing asymmetry of the hyperbolic layer for increasing values of $i_{0}$. At $i_{0}=65.2^{\circ}$, the hyperbolic layer is clearly more developed for the lower range of semi-major axis.
\end{enumerate}

The dynamics associated to the hyperbolic layer of
model $\mathcal{S}$ is similar to model $\mathcal{T}$ apart that the coefficients of the respective resonances are slowly modulated in time.  
This is exemplified for two orbits in the composite panel of Fig.\,\ref{fig:panel}, together with macro and micro views of the phase space structures.  
The orbit immersed within the stable region displays oscillations, whilst the orbit trapped into the hyperbolic layer displays the characteristic intermittency.  
Let us underline that the hyperbolic orbit, on the $20$ year timescale, displays U-turns (\ie the alternation between libration and circulation regimes of the resonant angle $u_{1}$) always directed towards lower semi-major axis, with a timescale of about $1.5$ year. The full homoclinic loop takes about $3$ years. 
We integrated the same orbit on a time interval $10$ times larger and we noticed  the unevenly  distribution between upper and lower U-turns,  the latter being more frequent.
This property is clearly inferred from the  thorough inspection and detailed geometry of the hyperbolic foldings near the saddle-like structure. The close-up view of the FLI map (see the magnified region materialised by the green box,  Fig.\,\ref{fig:panel}) reveals more foldings in the lower part of the chart. Increasing the parametric value of $i_{0}$ makes this property even sharper, as shown in the maps provided in Appendix \ref{app:maps}. The asymmetry of the foldings emerges from the fact that
\begin{align}
\delta(i_{0}) = \frac{\vert h_{2,0}(a,e,i)\vert}{\vert h_{2,2}(a,e,i)\vert},
\end{align} 
becomes larger along a solution 
$\bold{x}(t)=
\big(a(t),e(t),i(t))
\big)$ for initial conditions near $\bold{x}_{u}$.

The layer's dependence upon $\omega_{0}$,   for model $\mathcal{S}$, is shown for the  fixed value of $u_{1}=3.6$ in the last map of the composite plot of Fig.\,\ref{fig:panel} (bottom right). It reveals a much wider width (roughly speaking on the order of $10$ km in the semi-major axis) for $\omega \in [\pi,3\pi/2]$, with  petals structures. 
For $\omega \in [3\pi/2,\pi]$,  
the hyperbolic structure is much simpler to apprehend. 
For Molniya's typical variation of 
$\omega \sim 270^{\circ}\pm 20^{\circ}$, this corresponds to the rectangle 
materialized with white dashed lines.

\begin{figure}
	\centering
	\includegraphics[width=1\textwidth]{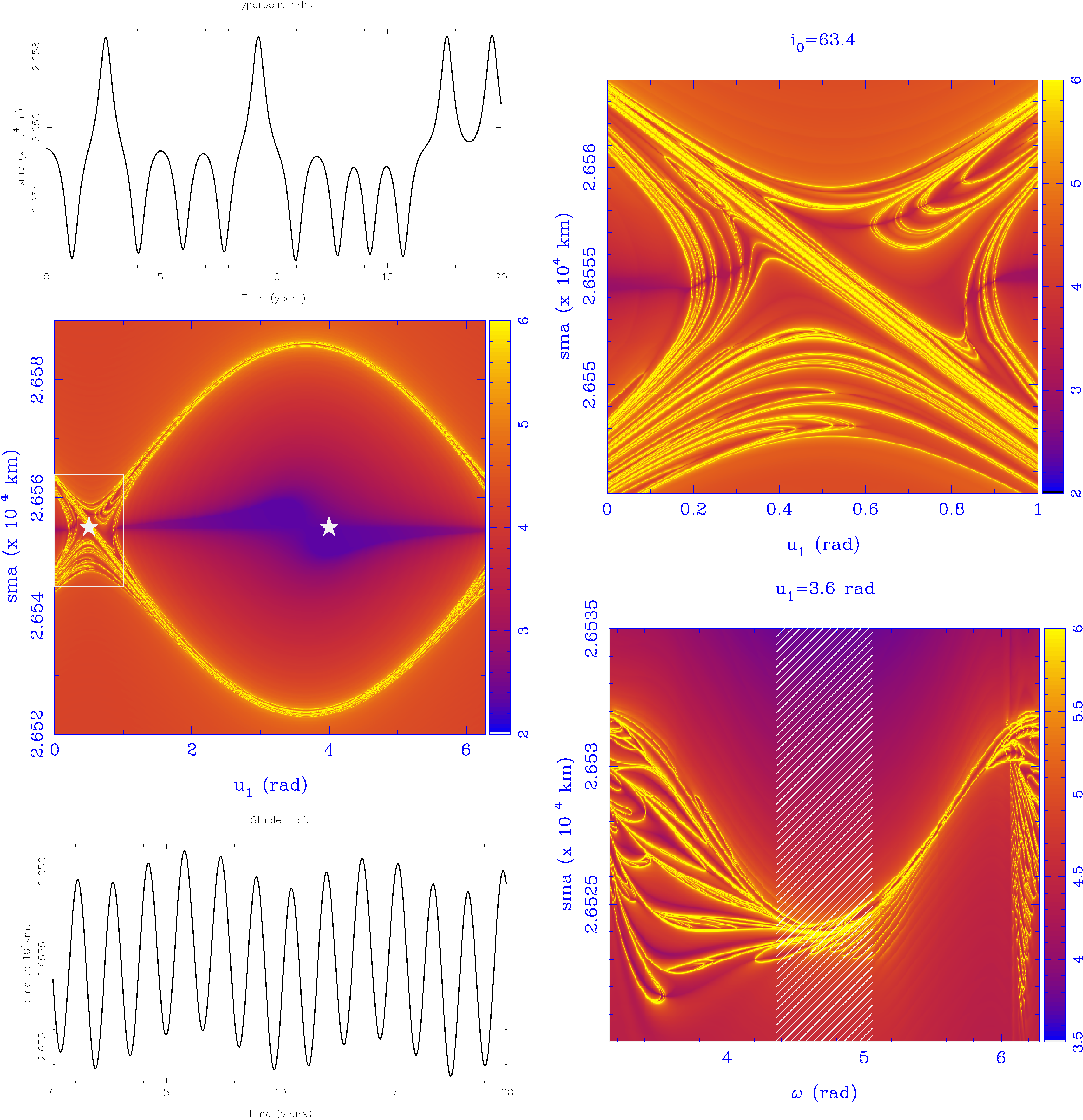}
	\caption{\label{fig:panel}
		Composite plot highlighting the main features of Molniya semi-major axis dynamics. The global FLI map and a magnified portion near the saddle-like structure detail the hyperbolic structure. Initial conditions within the hyperbolic layer display intermittency phenomena, whilst  stable  orbits display regular oscillations. This is exemplified for two orbits whose initial conditions are labeled with the white stars. 
		The width of the layer, for a fixed $u_{1}$ but varying $\omega$, might exhibit a complex geometry. For Molniya's prototypical  range of values of $\omega$, materialised by the white shaded-line region around $\omega=270^{\circ}$, the width is limited to a few kilometer in the semi-major axis only. See text for details. 	
	}
	\label{fig:panel}
\end{figure}

% ====================================
\section{Connections and links with the dynamics of Molniya 1-69 and Molniya 1-87}\label{sec:TLE}
\begin{figure}
	\centering
	\includegraphics[width=0.8\linewidth]{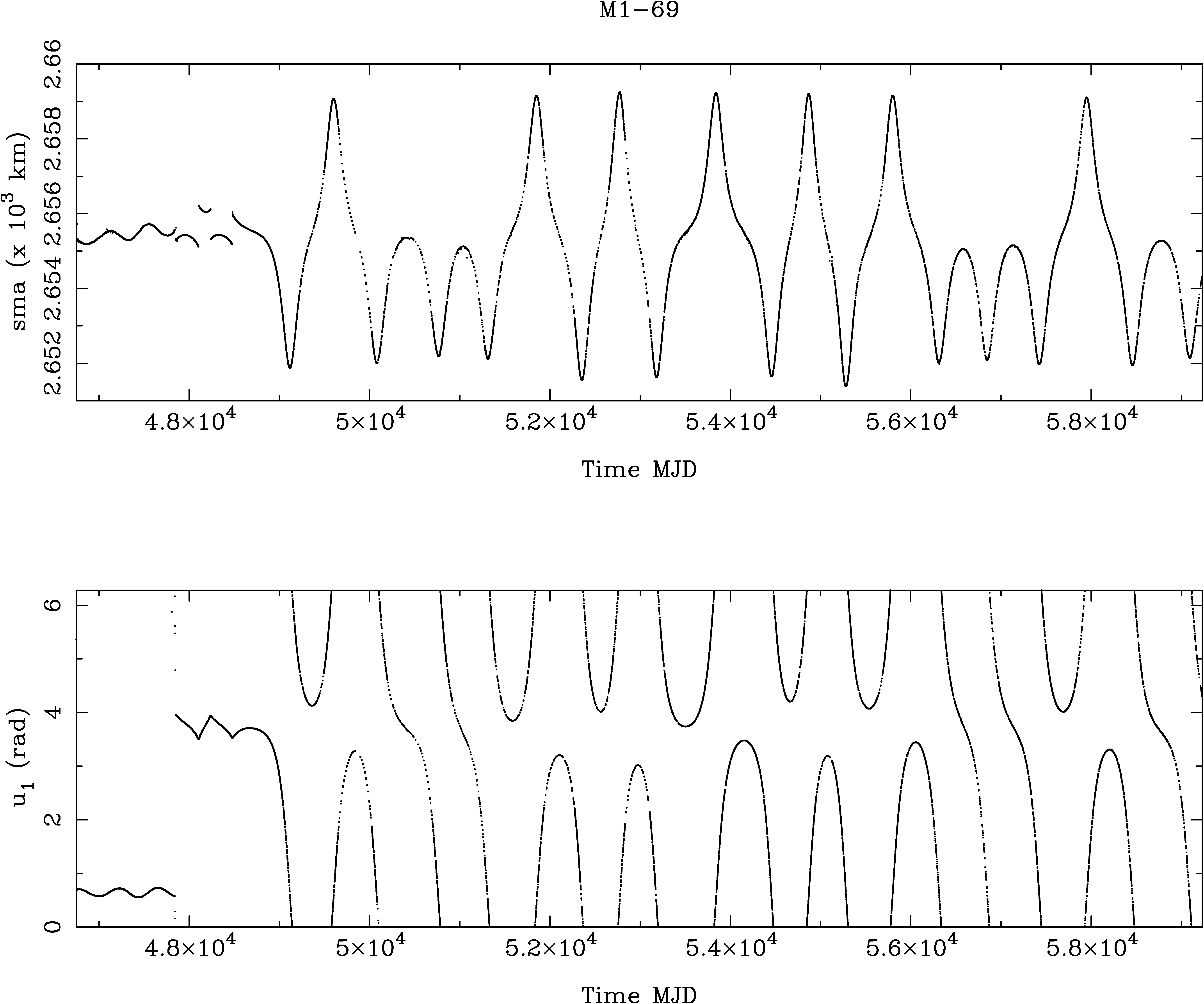}
	\caption{Time history of the semi-major axis 
		and resonant angle $u_{1}$ 
		extracted from the  TLE data for the satellite  Molniya 1-69.}
	\label{fig:tle69}
\end{figure}

\begin{figure}
	\centering
	\includegraphics[width=0.8\linewidth]{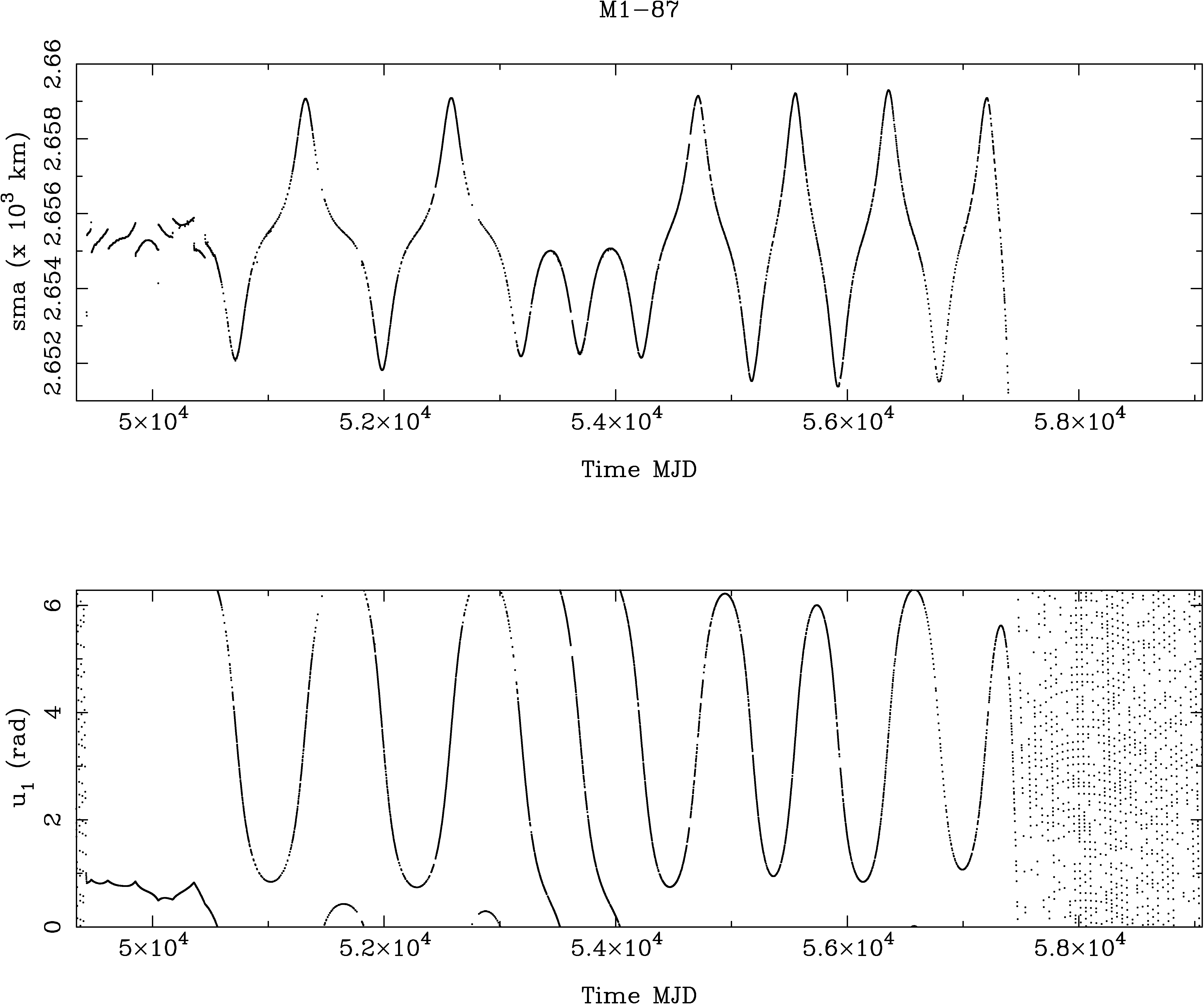}
	\caption{Time history of the semi-major axis 
		and resonant angle $u_{1}$ 
		extracted from the  TLE data for the satellite  Molniya 1-87}
	\label{fig:tle87}
\end{figure}

\begin{figure}
	\centering
	\includegraphics[width=1\linewidth]{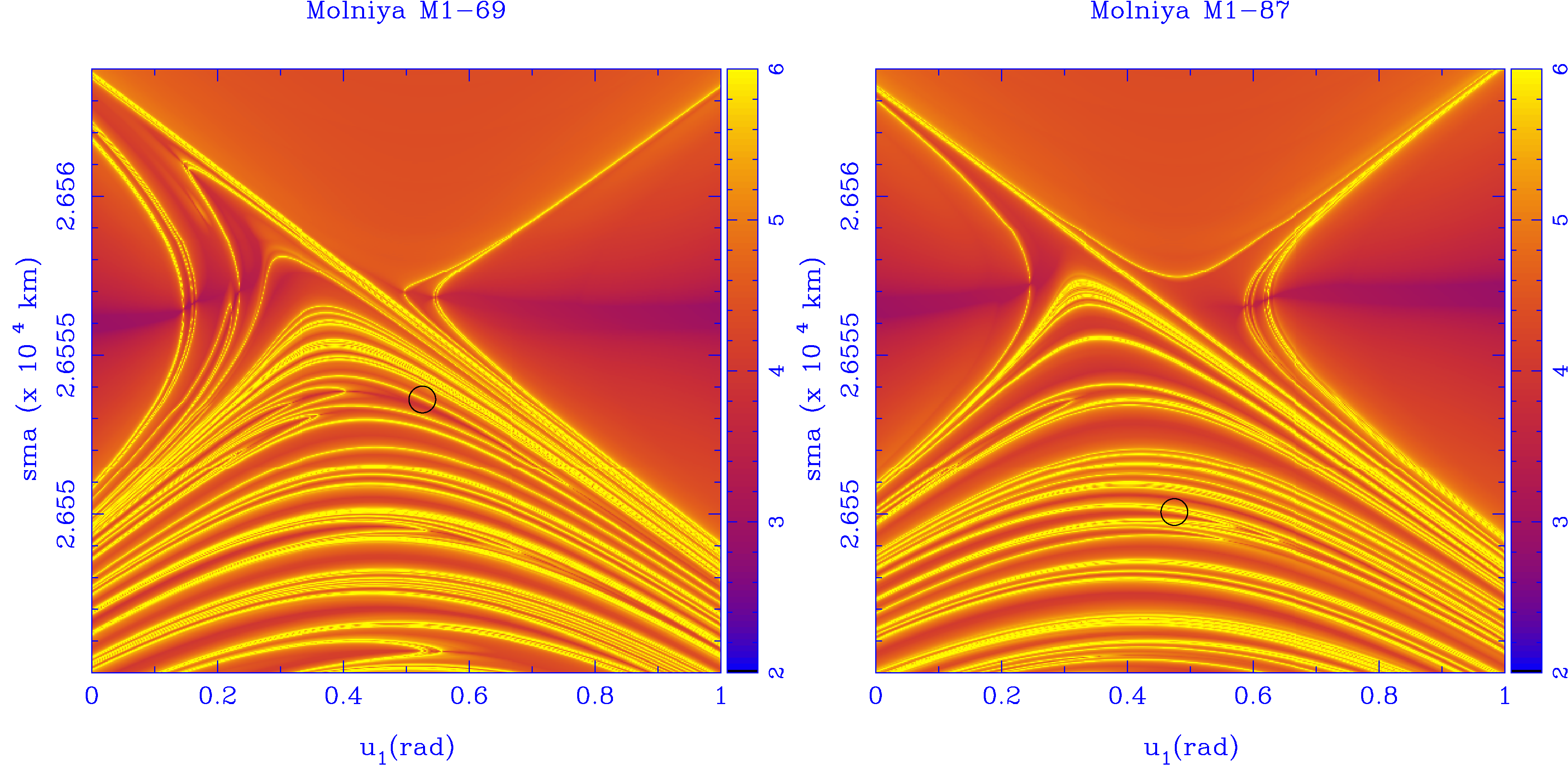}
	\caption{Dynamical maps for Molniya 1-69 and Molniya 1-87. The locations of Molniya 1-69 and Molniya 1-87 are marked through the black circle. Both satellites reside within the hyperbolic tangle.}
	\label{fig:TLEMAP}
\end{figure}

% ====================================
On inspecting the extracted semi-major axis using 
Molniya 1-69 and Molniya 1-87 TLE data, we notice that they display intermittency phenomena on their semi-major axis\footnote{
	We extracted the mean semi-major axis (in the sense of the underlying \texttt{SGP4} theory) from the TLEs by following the ``un-Kozai'' mean-motion procedure (one step iterative method) presented in \cite{fHo80}, section $6$. See also \cite{dVa08}, Eq. (7) or  \cite{fHo04}, appendix B, section A.} 
as repeated  in  Fig.\,\ref{fig:tle69} and Fig.\,\ref{fig:tle87}. The figures also incorporate the time evolution of the resonant angle $u_{1}$. The relevant part of the data, in the light of the oscillating models previously derived,  cover in both cases at least $2$ decades. Both data contain a transitory period, possibly remnants of unknown manoeuvres. For Molniya 1-87, after the epoch corresponding to mean Julian day (MJD) of $5.7 \times 10^{4}$, the satellite experienced a significant semi-major axis reduction. We will not pay attention  to this part of the data. In the exploitable window, the resonant signature, consisting of  alternation between libration and circulation,  is well-marked and in accordance with the U-turns intermittent semi-major axis variations. 
In both cases, the intermittency U-turns  take place for  $a \sim 26,550$ km, compatible  with the locations of the hyperbolic foldings we located with model $\mathcal{S}$ close to the saddle. 
It is worth mentioning that Molniya 1-69 has been left untouched in \cite{tlZh15}, as judged to ``locate in the vicinity of this separatrix''. Below, we give more credit to this claim,  and we show that it is also the case for its cousin Molniya 1-87. 
At the light of the dynamical mechanisms presented in section \ref{sec:LS} and the fingerprints just described, it is tempting to say that both satellites evolve within the hyperbolic layer. To give more weight to this claim, we performed the following steps: 
\begin{enumerate}
	\item At epochs $t_{\star}$ corresponding to the apex of the first U-turns, we extract from the TLEs the corresponding orbital parameters and we record the values of $(a^{\star},u_{1}^{\star})$. For case Molniya 1-69, we selected $t_{\star}=50,418.06$ (MJD), leading to the ``instantaneous'' elements
	\begin{align}
	\left\{
	\begin{aligned}
	& a=26,553.63 \, \textrm{km}, 
	\hspace{0.44cm} u_{1}=0.5257, \\
	& e=0.67633, \hspace{1.44cm} \omega=269^{\circ}95, \\
	& i=64^{\circ}2544, \hspace{1.4cm} \Omega=249^{\circ}68.
	\end{aligned}
	\right.
	\end{align} 
	For Molniya 1-87, we selected $t_{\star}=53,433.24$ (MJD), for which the sets of computed elements reads  
	\begin{align}
	\left\{
	\begin{aligned}
	& a=26,550.06 \, \textrm{km}, 
	\hspace{0.44cm} u_{1}=0.4749, \\
	& e=0.6582, \hspace{1.58cm} \omega=262^{\circ}68, \\
	& i=64^{\circ}1995, \hspace{1.4cm} \Omega=223^{\circ}01.
	\end{aligned}
	\right.
	\end{align}  
	\item We compute  the  dynamical maps
	with the FLIs for the vector field $v_{\mathcal{S}}$, using as parameters and phasing for the section
	$\Sigma$ those extracted from the respective TLE at epoch $t_{\star}$.
	\item On the obtained dynamical maps, the points of coordinates $\big(a(t_{\star}),u_{1}(t_{\star})\big)$ are spotted.   
\end{enumerate}
The obtained dynamics maps shown in Fig.\,\ref{fig:TLEMAP} convincingly demonstrate that the satellites reside within the hyperbolic tangle.

%===========================
\section{Conclusions}
%===========================
The constructed dynamical models  and their analysis allowed us to deepen the understanding
of Molniya's semi-major axis dynamics.  
The hyperbolic structures organising the phase space 
have been portrayed via variational indicators through a series of compact, tractable and realistic secular  models. The effect of lunisolar perturbations, on the $20$ years timescale, needs to be taken into account to reconstruct the correct dynamical template. 
In fact, the induced modulations of the eccentricity and inclination contribute sufficiently  to change  the ``parameters'' of the tesseral problem; the coefficients we denoted by $h_{2,0}$ and $h_{2,\pm 2}$. 
We connected the $20$ year long fingerprints  of two satellites, Molniya 1-69 and Molniya 1-87,   with the hyperbolic layer surrounding the unperturbed cat-eye separatrix. This hyperbolic layer, in absence of lunisolar perturbations, would be too thin to sustain the dynamical signatures visible at the publicly available data level. 
By computing their associated dynamical maps, 
we provided evidence that the two satellites are  trapped within the hyperbolic tangle. The secular 
dynamics umbrella provided a reliable and robust mold
to approach and explain the  semi-major axis patterns extracted from the TLE space datasets.
As far as we are aware, this result is 
the first report of long time scale hyperbolicity corroborated by pseudo-observations in the near-Earth space environment. The mechanisms and tools depicted in this contribution have relevance for other dynamical regions, most notably for the geosynchronous altitude where 
similar patterns have been observed on simulated orbits \cite{sBr05,iWy07,sPr21}.

\section*{Acknowledgments} 
J.D. is a postdoctoral researcher of the ``Fonds de la Recherche Scientifique'' - FNRS. 
J.D. acknowledges discussions with Florent Deleflie, Denis Hautesseres, Alexis Petit 
and David Vallado about the TLEs data, and discussions that have followed from the CNES \textsc{COMET-ORB} workshop on ``Uncertainty Quantification in Orbit Propagation'', Feb. $9-10$, $2021$. J.D. and A.L. acknowledge several discussions with Alessandra Celletti and C\u{a}t\u{a}lin Gale\c{s} on the resonant potential.
J.D. acknowledges discussions with Ioannis Gkolias on the $J_{2}^{2}$ effect and useful references provided. J.D. acknowledges several discussions all along this research with  Aaron Rosengren. 

\section*{Conflict of interest}
The authors declare that they have no conflict of 
interest.

\newpage
\bibliographystyle{plain}      % basic style, author-year citations
\bibliography{biblio}

\appendix
\section{Dynamical models employed \& physical parameters}\label{app:ForceModel}
As we mentioned in the introduction, the aim of this study is not to study with the greatest accuracy possible Molniya dynamics, with a comprehensive force model including uncertainty modeling and Monte Carlo like approaches. Quite on the contrary, we leverage the understanding of the dynamics of the semi-major axis from the essential ``building blocks'' with tractable contributions. In that respect, we would like to 
provide more context to the dynamical model we have employed. We have approached the problem as a drag-free model, with no solar radiation pressure,  based on a compact geopotential model including relevant terms of the disturbing lunisolar potentials. 
Higher order zonal secular terms can be obtained in closed form over the eccentricity following the
same formal procedure  discussed in Section \ref{subsec:zonal}. In terms of the orbital elements, up to order $l=5$, they read:
\begin{align}
\left\{
\begin{aligned}
&V_{J_{3}}= 
\frac{3J_{3}\mu r_{\Earth}^{3}e \sin i (5 \sin^{2}i-4)}
{8a^{4}(1-e^{2})^{5/2}} \sin(\omega), \\
&V_{J_{4}}= \Big(
\frac{15 J_{4} \mu r_{\Earth}^{4} \sin^{2}i(7\sin^{2}i-6)}
{64 a^{5}(1-e^{2})^{7/2}}
\Big)
\cos(2\omega) -
\frac{3J_{4}\mu r_{\Earth}^{4}(3e^{2}+2)(35\sin^{4}i-40 \sin^{2}i+8)}{128 a^{5}(1-e^{2})^{7/2}}
, \\
&V_{J_{5}}= 
\Big(
\frac{15J_{5}\mu r_{\Earth}^{5} e(3e^{2}+4)\sin i (21 \sin^{4}i -28 \sin^{2}i +8)}
{128 a^{6}(1-e^{2})^{9/2}}
\Big)
\sin(\omega)
-
\Big(
\frac{35J_{5}\mu r_{\Earth}^{5} e^{3}\sin^{3}i (9\sin^{2}i-8)}
{256 a^{6}(1-e^{2})^{9/2}}
\Big)
\sin(3\omega) \notag.
\end{aligned}
\right.
\end{align} 
It is worthwhile to note that the resonant argument of perigee also appears in the above secular contributions; hence the idea that Molniya orbits, besides tesseral and lunisolar resonances, gather also ``zonal resonances''. 
To include the second-order part term with factor $J_{2}^{2}$ in the secular Hamiltonian, with the form 
\begin{align}
V_{J_{2}^{2}} = J_{2}^{2}
\big(
A(a,e,i) \cos 2 \omega + B(a,e,i)
\big)
,
\end{align} 
we used the formula given in \cite{sBr98,mLa14}.  The relevance of our model $\mathcal{S}$ has been assessed by including those effects, and the lunisolar 
$h_{h}^{\Moon\Sun} \cos h$
to $\mathcal{L}$. This model forms an ``extended'' Hamiltonian model  $\tilde{\mathcal{S}}$. 
We computed the dynamical map for the Hamiltonian vector field  $v_{\tilde{\mathcal{S}}}$ with $e=0.7$, $i_{0}=64^{\circ}3$ and $(\omega,\Omega)=(270^{\circ},0)$ and we did not noticed significant macroscopic changes in the obtained dynamical template;  henceforth the relevance of the Hamiltonian model $\mathcal{S}$. Let us mention that even if the macroscopic structures do not change drastically, hyperbolic orbits 
generated under model $\mathcal{S}$ and $\mathcal{\tilde{S}}$ will separate in time (sensitivity to the slight change of physics), and the hope to follow them beyond a few Lyapunov times is a useless effort. The Lyapunov time $\tau_{\mathcal{L}}$ computed as 
\begin{align}
\left\{
\begin{aligned}
&\chi = \lim_{s \to +\infty} \frac{1}{s} \log(
\norm{w(s)}), \, \notag \\
& \tau_{\mathcal{L}} = 1/\chi,
\end{aligned}	
\right.
\end{align}    
is about $2$ decades.     \\

The physical parameters of this study read as follow. 	
The Moon's orbital parameters, referred to the ecliptic plane, have been set to $a_{\Moon}=384,748$ km, $e_{\Moon}=0.0554$, $i_{\Moon}=5^{\circ}15$, $\mu_{\Moon}=4902.8$ $\textrm{km}^{3}/\textrm{s}^{2}$. 
The Sun's orbital parameters, referred to the Earth equator, have been set to  $a_{\Sun}=1.496 \times 10^{8}$ km, $e_{\Sun}=0.0167$, $i_{\Sun}=23^{\circ}4392911$, $\mu_{\Sun}=1.32712 \times 10^{11}$ $\textrm{km}^{3}/\textrm{s}^{2}$ . The length unit is the Earth radius $r_{\Earth}$  of $6378.1363$ km, $\mu=398,600.44$ $\textrm{km}^{3}/\textrm{s}^{2}$.

\section{Dynamical maps}\label{app:maps}
\begin{figure}
	\centering
	\includegraphics[width=0.9\textwidth]{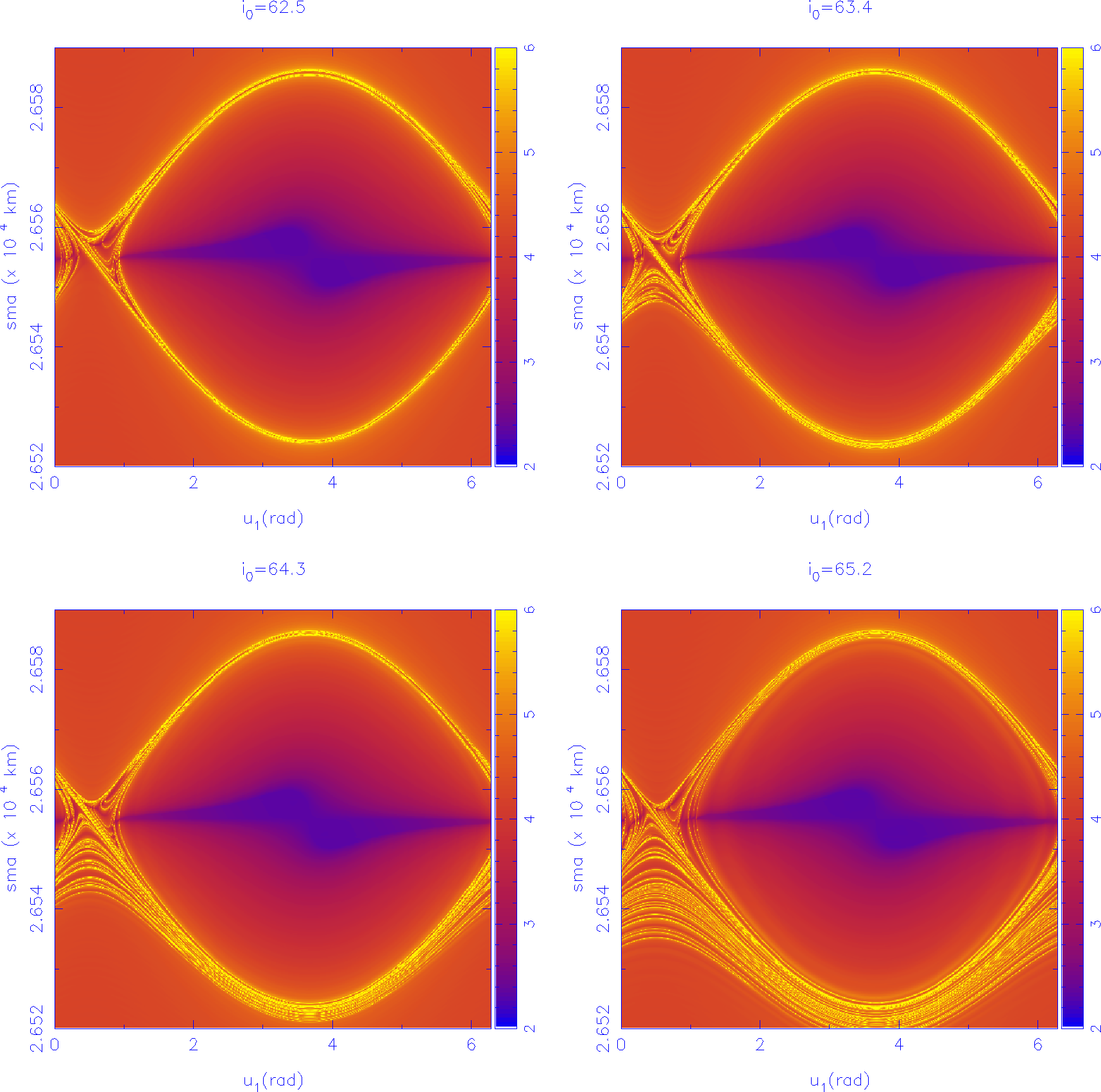}
	\caption{Intersections of the forward in time FLIs  
		with the plane $(a,u_{1})$ for model $\mathcal{S}$ computed on a $500 \times 500$ grid of initial conditions for $i_{0} \in \{62.5, 63.4, 64.3, 65.2\}$ deg.  	
	}
	\label{fig:modelS}
\end{figure}

\begin{figure}
	\centering
	\includegraphics[width=0.9\textwidth]{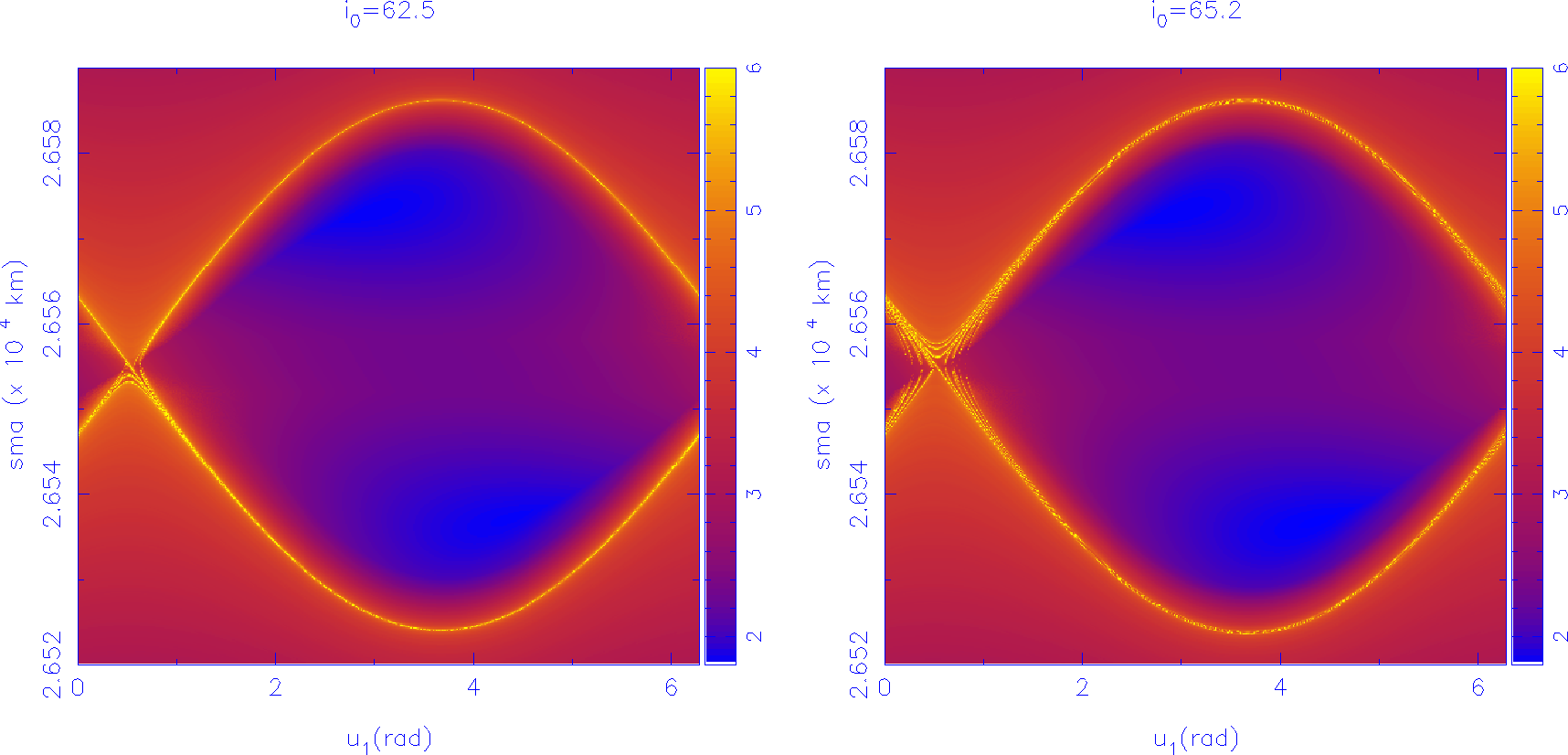}
	\caption{Intersections of the forward in time FLIs  
		with the plane $(a,u_{1})$ for model $\mathcal{J}$ computed on a $500 \times 500$ grid of initial conditions for $i_{0} \in \{62.5,65.2\}$ deg.  	
	}
	\label{fig:modelJ}
\end{figure}

We computed dynamical maps  for a fixed value of  $e=0.7$ and  $i_{0}$ ``piercing'' the critical inclination. They are presented in  Fig.\,\ref{fig:modelS} for model $\mathcal{S}$. Given that model $\mathcal{J}$ is slightly perturbed, we just show the maps for 
$i_{0}=62^{\circ}5$ and  $i_{0}=65^{\circ}2$ in Fig.\,\ref{fig:modelJ}.
The maps have been computed on a $500 \times 500$ grid of initial conditions, forward in time, and over a time interval of $20$ years. We have considered $4$ values of the initial inclinations, namely 
$i_{0} \in \{62^{\circ}5, 63^{\circ}4, 64^{\circ}3, 65^{\circ}2\}$. The initial phasing is set as $(\omega,\Omega)=(270^{\circ},0^{\circ})$. If a given initial condition in the map fall within the highest region of the FLIs (yellow tone), then the orbit is hyperbolic and exhibit sensitive dependence upon the initial condition (\ie any orbit starting with an initial condition slightly different will have a long-term different future; the orbits will separate with time). We note 
that the $i_{0}$-dependence of the $\mathcal{J}$ model is quasi-absent. The model $\mathcal{J}$ is very close to the integrable picture, in the sense that the splitting of the separatrix is weak.  The latter is much more manifest for model $\mathcal{S}$, where we recall, the eccentricity and inclinations variables are no longer frozen. For increasing values of $i_{0}$, we underline the growing asymmetry of the foldings near the saddle-like structure for the model $\mathcal{S}$. This particular structure transfers directly at the single orbit level: an orbit trapped  within the hyperbolic layer is more likely to display U-turns intermittency phenomena towards the lower semi-major axis. This observation, based on the thin structures of the lobes detected with a variational dynamical indicator on our model, is also in agreement with the actual two-line elements datasets for objects M1-69 and M1-87.    

\end{document}